\documentclass[aps,prx,reprint,twocolumn,superscriptaddress,floatfix,nofootinbib,longbibliography]{revtex4-2}
\usepackage{graphicx,amsmath,amsfonts,amssymb,amsthm,xr}
\usepackage{epsfig,amsmath,amssymb,color,dsfont,upgreek,physics}
\usepackage{mathrsfs}
\usepackage{mathtools}
\usepackage{bbold}
\usepackage{float}
\usepackage[caption=false]{subfig}
\usepackage[bookmarks=true,colorlinks,linkcolor=OrangeRed,urlcolor=NavyBlue,citecolor=RoyalBlue]{hyperref}
\usepackage[capitalize]{cleveref}
\usepackage[dvipsnames]{xcolor}
\usepackage{orcidlink}

\usepackage{graphicx}
\usepackage{dcolumn}
\usepackage{bm}
\usepackage{color}
\newcommand{\Zt}{\mathbb{Z}_2}
\newcommand{\rr}{\mathbf{r}}

\newcommand{\tightmathspacing}{%
  \setlength{\parskip}{0pt}%
  \setlength{\abovedisplayskip}{6pt plus 2pt minus 2pt}%
  \setlength{\belowdisplayskip}{6pt plus 2pt minus 2pt}%
  \setlength{\abovedisplayshortskip}{4pt plus 2pt minus 1pt}%
  \setlength{\belowdisplayshortskip}{4pt plus 2pt minus 1pt}%
}

\tightmathspacing

\begin{document}

\title{Statistics as a local phase: crystalline order and quench dynamics of emergent dimers in Ising gauge theories}

\author{Umberto Borla${}^{\orcidlink{0000-0002-4224-5335}}$}
\email{umberto.borla@lmu.de}
\affiliation{Max Planck Institute of Quantum Optics, 85748 Garching, Germany}
\affiliation{Department of Physics and Arnold Sommerfeld Center for Theoretical Physics (ASC), Ludwig Maximilian University of Munich, 80333 Munich, Germany}
\affiliation{Munich Center for Quantum Science and Technology (MCQST), 80799 Munich, Germany}

\author{Riccardo Cioli}
\affiliation{Dipartimento di Fisica e Astronomia, Università di Bologna, I-40127 Bologna, Italy}

\author{Jad C.~Halimeh${}^{\orcidlink{0000-0002-0659-7990}}$}
\email{jad.halimeh@lmu.de}
\affiliation{Department of Physics and Arnold Sommerfeld Center for Theoretical Physics (ASC), Ludwig Maximilian University of Munich, 80333 Munich, Germany}
\affiliation{Max Planck Institute of Quantum Optics, 85748 Garching, Germany}
\affiliation{Munich Center for Quantum Science and Technology (MCQST), 80799 Munich, Germany}
\affiliation{Department of Physics, College of Science, Kyung Hee University, Seoul 02447, Republic of Korea}

\date{\today}

\begin{abstract}
    How does the Bose or Fermi statistics of microscopic particles survive when confinement binds them into emergent bosonic composites? We address this question in the strong-coupling limit of a $2+1$D $\mathbb{Z}_2$ lattice gauge theory, where charges are confined into tightly bound pairs that can be described by an effective dimer model. We find that the statistics of the underlying matter is encoded entirely in a single local hopping phase $\varphi$---$0$ for bosons, $\pi$ for fermions. Treating $\varphi$ as a continuous parameter that interpolates between the two, we map the ground-state phase diagram with the help of tensor-network methods. The angle $\varphi$ itself drives a transition between a dimer-superfluid and a dimer-charge density wave state, while the magnetic coupling binds neighboring dimers into resonating pairs, in competition with the inter-dimer repulsion. We identify a novel gapped phase in which dimer pairs crystallize into an ordered pattern of resonating plaquettes. Finally, we propose a quench protocol under which identical dimer configurations evolve in markedly different ways depending on the statistics of their constituents. This provides a dynamical probe of the internal structure of dimers, and allows to detect ordered phases through real-time signatures, within reach of simulators that natively realize bosonic degrees of freedom.
\end{abstract}

\maketitle

\section{Introduction}
\label{sec:intro}
Dimer models play a central role in the understanding of strongly correlated quantum matter. Originally introduced in the context of resonating valence bond physics and high-temperature superconductivity \cite{anderson1987, kivelson1987, rokhsar1988}, they provide minimal effective descriptions of systems where local constraints and short-range pairing dominate the low-energy behavior. Despite their formal simplicity, dimer models host remarkable phenomenology, including crystalline ordered phases \cite{Chayes1989, sachdev1989, chaubey2026quantumdimerspifluxtoric}, topological liquids \cite{moessner2001rvb, moessner2003} and exotic critical behavior \cite{rokhsar1988, ARDONNE2004493, Sachdev2008, Sachdev_2011}. 
 \begin{figure*}[t]
     \centering
     \includegraphics[width=0.95\linewidth]{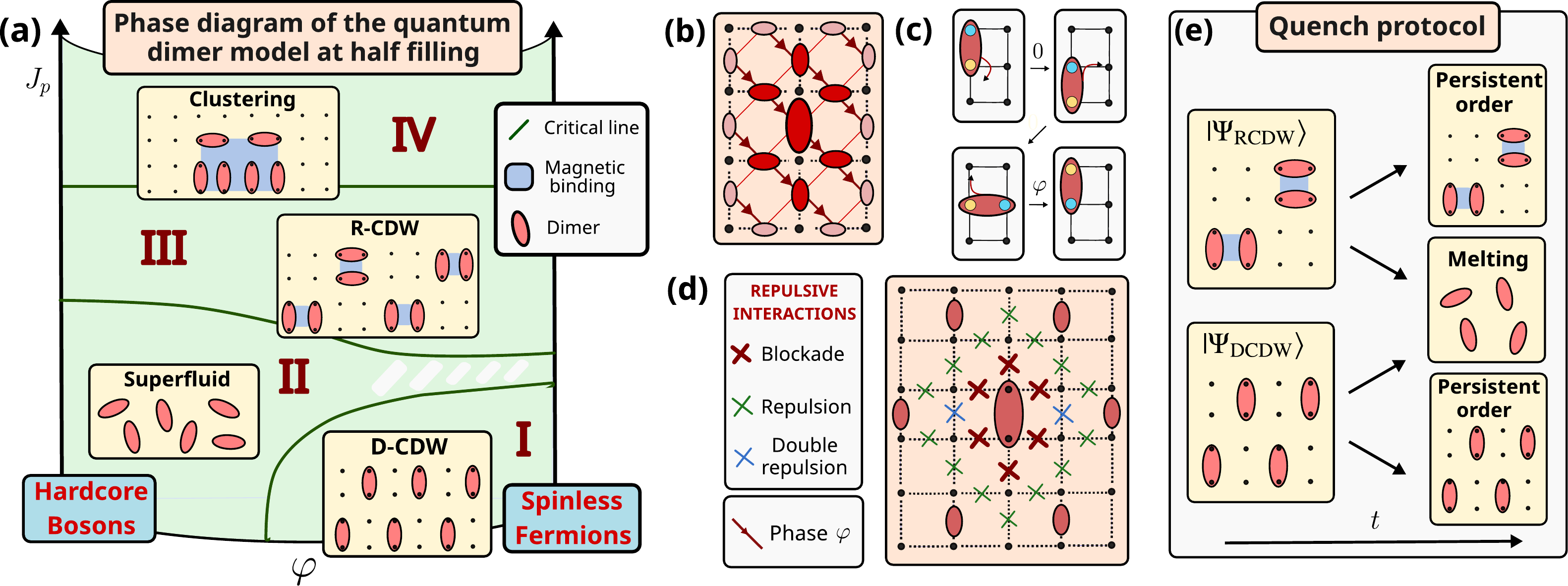}
     \caption{\textbf{(a)}: Ground state phase diagram of the dimer model describing the strong coupling regime of a $\Zt$ LGT with bosonic (fermionic) matter at $\varphi=0$ ($\pi$). The angle $\varphi$ interpolates between the two, driving a transition between a dimer superfluid (II) and a dimer charge density wave (I) at zero magnetic coupling $J_p$. For $J_p>0$ dimers bind into resonating pairs, which form a new ordered ``resonating charge density wave'' phase (III). At large $J_p$ the magnetic pairing causes all dimers to cluster together, leading to an additional phase (IV). \textbf{(b)}: Depiction of the original lattice (black vertices) and of the dual lattice obtained by connecting the centers of its links. The central dimer occupying a vertex of the dual lattice can hop to the highlighted positions, gaining a phase $\varphi$ when following one of the arrows. \textbf{(c)}: sequence of hoppings resulting in a change in the orientation of the dimer. The state picks up a phase $\varphi$ due to the phase pattern in \cref{eq:qdh}. \textbf{(d)}: Sketch of the D-CDW state and of the repulsive interactions of the effective dimer model \cref{eq:eff_repulsion}. \textbf{(e)}: Sketch of the dynamical protocol described in \cref{sec:dyn}. The stability of initial ordered states is probed through quenches to different parameter regimes.}
     \label{fig:phase_diagram}
 \end{figure*}
While dimer models are often introduced phenomenologically, an important question concerns their emergence from microscopic Hamiltonians \cite{Fradkin_2013_dimer}. In the most conventional approach, dimers are identified with fluctuating spin-singlet bonds in quantum magnets \cite{anderson1987, rokhsar1988, moessner2011} and, more recently, have been connected to certain phases of Rydberg array systems  \cite{samajdar2021rydberg, verresen2021prediction, zeng2025dimer}. A related but conceptually distinct route emerges from lattice gauge theories with dynamical matter, where confinement naturally gives rise to tightly bound charge pairs \cite{ROSSI1984, moessner2001ising, borla2020confined, borla2022, borla2026odd, Sachdev_2023}. In this regard, a question that remains largely unexplored is how the statistics of microscopic constituents affects the emergent dimer model description. While models including both fermionic and bosonic dimers ---of distinct microscopic origin--- have been studied \cite{pollmann2011fermionic, punk2015, feldmeier2018exact}, it is not clear how dimers of identical statistics are affected by their internal composition, and how this is encoded in the emergent description. 

This article focuses on dimer models at finite filling, emerging as strong-coupling descriptions of $\Zt$ lattice gauge theories with dynamical matter \cite{wegner1971, fs1979, senthil2000, assaad2016, gazit2017, kebric2021, borla2022, borla2024, borla2026odd, kebric2026} in two spatial dimensions. In the specific particle-number-conserving model that we consider, the ground state properties are known to depend strikingly on the statistics of the matter fields. Hardcore bosons realize a superfluid phase of confined dimers (D-SF) \cite{Homeier2023, kebric2026}, while fermionic matter stabilizes a gapped dimer charge density wave (D-CDW) state at half filling \cite{borla2022, borla2024}. We show that within the effective model the distinction between the two is encoded exclusively through a characteristic pattern of hopping phases, associated with changes in dimer orientation. This perspective suggests a continuous interpolation between bosonic and fermionic constituents through an effective ``statistical angle'' $\varphi$, prompting us to study how the ground state and non-equilibrium dynamics evolve between the two limits. A continuous transition between the D-SF and D-CDW phases takes place at finite $\varphi$ when a plaquette term is absent, while its presence mediates a form of pairing between the dimers. The regime where such a term competes with the inter-dimer repulsion is characterized by a possibly novel ordered state, where emergent magnetically bound degrees of freedom arrange themselves into a crystalline pattern with wave vector $\mathbf{k}=(\pi/2, \pi/2)$. This reveals an intriguing hierarchy of composite objects, which arise naturally from the microscopic model without requiring fine tuning or \textit{ad hoc} interactions: electric confinement leads to emergent $\Zt$-neutral dimers, which in turn are bound into resonating pairs by the magnetic term. Both objects can form gapped ordered states stabilized by the emergent effective repulsion. Transitions between the aforementioned quantum phases involve an interplay of magnetic binding and different kinds of spatial ordering, and are potentially described by quantum critical points which fall beyond Landau's symmetry-breaking paradigm \cite{senthil2004, senthil2004review, levin2004, Sachdev2008, wang2017deconfined, senthil2023deconfinedquantumcriticalpoints}. 

$\Zt$ lattice gauge theories have drawn considerable interest in recent times in the context of quantum simulation \cite{aidelsburger2021, zohar2021, Klco_2022, Irmejs2023minimal, Bauer2023, bauer2023hep, dimeglio2024, mildenberger2025, Halimeh2025, halimeh2025quantumsimulationoutofequilibriumdynamics}, as a paradigmatic example of models which exhibit non-trivial phenomenology while maintaining a simple local Hilbert space \cite{zohar2017digital, barbiero2019coupling, lumia2022, Homeier2023}. The latter feature makes them strong candidates for experiments on current quantum devices, which have so far focused on high energy physics phenomenology such as string dynamics \cite{alexandrou2025, Cochran2025, borla2025stringbreaking, cobos2025realtimedynamics21dgauge, Xu2025StringBreakingGlueball, xu2026observationglueballexcitationsstring} and disorder-free localization \cite{googledfl2026}. A complementary direction consists in probing properties of strongly correlated systems, including ordered phases and non-conventional phase transitions \cite{homeier2021, Homeier2023}. A direct implementation of the effective dimer models on quantum hardware would allow to inspect real-time signatures of the aforementioned ordered phases in highly entangled dynamical regimes, which are difficult to access with classical methods. To this end, we devise a protocol to probe the quench dynamics of simple ordered initial states. Tensor network simulations show how identical dimer configurations can exhibit markedly distinct evolution depending on the microscopic statistics of their constituents, ranging from rapid relaxation to persistent ordering. In this way, we indirectly probe the internal structure of the dimers and how it affects their real-time evolution. Since the effective description encodes the underlying fermionic statistics in a local phase, our quench protocol is accessible even to platforms that natively realize bosonic degrees of freedom, without the complications that typically arise from encoding fermionic matter \cite{BRAVYI2002210, derby2021}. 

This article is structured as follows: In \cref{sec:model} we introduce the $\Zt$ LGT Hamiltonian, and show how its strong-coupling limit is described by an effective dimer model with repulsive and Rokhsar--Kivelson interactions. In \cref{sec:statistics} we discuss how the microscopic statistics is encoded in the effective model, and motivate an extension of the model to complex hopping phases. In \cref{sec:gs} we derive the full ground state phase diagram of the effective model, and discuss the nature of the quantum phase transitions that occur. In \cref{sec:dyn} we devise a quench protocol which, given simple spatially ordered initial states, tests the  stability of their order with respect to the parameters of the effective Hamiltonian. In \cref{sec:concl} we discuss results and possible extensions of our study.

\section{Model}
\label{sec:model}
We consider a $2+1$D $\Zt$ lattice gauge theory with Hamiltonian
\begin{equation}
\hat{H}=-J_p\sum_{\rr^*} \hat B_{\rr^*}
-\kappa \sum_{\rr,\eta} \big(\hat a^\dagger_\rr \hat\sigma^z_{\rr,\eta} \hat a_{\rr+\eta} +\text{H.c.}\big)-h\sum_{\rr,\eta} \hat\sigma^x_{\rr,\eta},
\label{eq:2DFS}
\end{equation}
where the matter fields $\hat a^\dagger_\rr$ defined on the sites of the square lattice can be either spinless fermions or hardcore bosons. In the gauge sector, $\hat\sigma^{x}_{\rr,\eta}$ denotes the $\Zt$ electric field operator on the link emanating from site $\rr$ in the direction $\eta$, $\hat\sigma^{z}_{\rr,\eta}$ is the minimal coupling, while $\hat B_{\rr^*}=\prod_{b\in \square_{\rr^*}} \hat\sigma^z_b$ is the four-body plaquette operator, with $\rr^*$ labeling the plaquette centers. We also define the star (or vertex) operator $\hat A_{\rr}=\prod_{\eta \in +_\rr}\hat\sigma^x_{\rr, \eta}$ as the product of $\hat\sigma^x$ over the four links meeting at the vertex $\rr$. The Hamiltonian has a global $U(1)$ symmetry associated with conservation of the total particle number $N=\sum_\rr \langle \hat n_\rr \rangle$, so that each filling sector can be studied independently. It is moreover invariant under the local gauge transformations $\hat{G}_\rr = \hat A_{\rr}(-1)^{\hat n_\rr}$,
which relate the $\Zt$ electric lines $\hat{\sigma}^x=-1$ emanating from a site to its total $\Zt$ charge. Physical states satisfy Gauss's law, $\hat{G}_\rr|\psi\rangle=Q_\rr |\psi \rangle$, where $Q_\rr=\pm 1$ denotes the absence or presence of a static background $\Zt$ charge on that site. It follows from Gauss's law that occupied matter sites must source electric lines, whose cost is proportional to their length $l$ and to the electric coupling $h$. 

We now zoom in on the strong-coupling limit $h \rightarrow \infty$, which can be described in terms of an effective dimer model. In this limit, transitions that change the total electric field are energetically suppressed, resulting in an emergent conservation law. For a given total particle number $N$, the lowest energy sector consists of pairs of particles (dimers) connected by electric strings of unit length. We denote the dimer number operator as $\hat n^d_{\rr,\eta}$, and their creation and annihilation operators as $d^\dagger_{\rr, \eta}$ and $d_{\rr, \eta}$ respectively. The dimers are bosonic objects living on the links of the original square lattice, which obey the hardcore constraint $(d^\dagger)^2=0$ and the extended hardcore constraint $d^\dagger_{\rr_1, \eta_1} d^\dagger_{\rr_2, \eta_2}=0$ whenever $(\rr_1, \eta_1)$ and $(\rr_2, \eta_2)$ are neighboring links. We note that the total number of dimers is simply related to the microscopic filling by $N_d=N/2$. 

In the absence of a plaquette term, the effective Hamiltonian can be obtained through second-order perturbation theory. Virtual processes that extend and then shrink individual dimers contribute both a hopping term and repulsive interactions. The hopping part of the effective model was derived in \cite{borla2022}, and is given by

\begin{eqnarray} \label{eq:qdh}
 &H_{\text{hop}} =-\kappa_\text{d}\sum \bigg( |\vcenter{\hbox{\includegraphics[height=0.005\textheight]{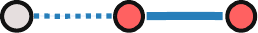}}}\rangle\langle\vcenter{\hbox{\includegraphics[height=0.005\textheight]{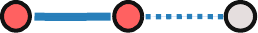}}}|+|\vcenter{\hbox{\includegraphics[height=0.025\textheight]{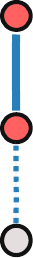}}}\rangle\langle\vcenter{\hbox{\includegraphics[height=0.025\textheight]{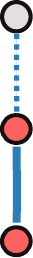}}}|\nonumber
+e^{i \varphi}|\vcenter{\hbox{\includegraphics[height=0.02\textheight]{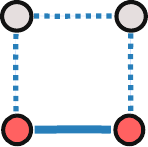}}}\rangle\langle\vcenter{\hbox{\includegraphics[height=0.02\textheight]{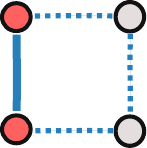}}}|\\
&+|\vcenter{\hbox{\includegraphics[height=0.02\textheight]{figures/term7.pdf}}}\rangle\langle\vcenter{\hbox{\includegraphics[height=0.02\textheight]{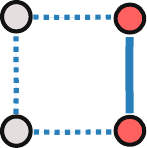}}}|
+ |\vcenter{\hbox{\includegraphics[height=0.02\textheight]{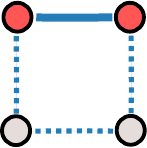}}}\rangle\langle\vcenter{\hbox{\includegraphics[height=0.02\textheight]{figures/term8.pdf}}}| +e^{-i \varphi}|\vcenter{\hbox{\includegraphics[height=0.02\textheight]{figures/term10.pdf}}}\rangle\langle\vcenter{\hbox{\includegraphics[height=0.02\textheight]{figures/term9.pdf}}}|+\text{h.c.}\bigg),
\end{eqnarray}
where $\kappa_{\text{d}}=\kappa^2/2h$ and the angle $\varphi$ is $0$ if the matter fields are bosonic and $\pi$ if they are fermionic. The single-particle band structure that follows from the hopping Hamiltonian \cref{eq:qdh} is discussed in \cref{app:single_dimer}.

Dimers $d_1^\dagger$ and $d_2^\dagger$ are defined to be neighbors if one (or both) sites covered by $d_2^\dagger$ can be reached from $d_1^\dagger$ by a single hopping of one of its constituents.  Repulsive interactions originate from the fact that when dimers are neighboring, certain energy-reducing virtual length-fluctuation processes are inhibited, penalizing such configurations. One has 
\begin{align}
    H_{\text{int}} = + U_{\text{d}} \smashoperator{\sum_{\mathbf{(r_1^*, r_2^*)}\in \mathcal{N}_1}}\,\,\hat{n}^d_{\mathbf{r_1*}}\hat{n}^d_{\mathbf{r_2*}}+ 2U_{\text{d}} \smashoperator{\sum_{\mathbf{(r_1^*, r_2^*)}\in \mathcal{N}_2}} \,\,\hat{n}^d_{\mathbf{r_1*}}\hat{n}^d_{\mathbf{r_2*}} 
    \label{eq:eff_repulsion}
\end{align}
where $\mathbf{r^*}$ identifies the links of the lattice, $U_{\text{d}}=\kappa^2/h=2\kappa_{\text{d}}$ and $\mathcal{N}_1$ ($\mathcal{N}_2$) are the sets of pairs of links with one (two) neighboring sites. The interactions are the same for fermions and bosons. A schematic illustration is provided in \cref{fig:phase_diagram}d. 

The plaquette term is easily included in the effective model, as in the unit-length dimers sector it can only act by flipping the orientation of two parallel dimers on the same plaquette:
\begin{equation} 
H_\text{plaq} =-J_p\sum_\square \hat R_\square, \qquad \hat R_\square=\left(|\vcenter{\hbox{\includegraphics[height=0.02\textheight]{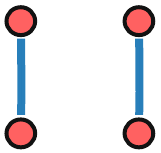}}}\rangle\langle\vcenter{\hbox{\includegraphics[height=0.02\textheight]{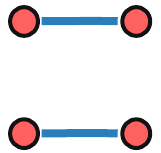}}}|+\text{h.c.}\right).
\label{eq:RKN}
\end{equation}
This is nothing but the standard kinetic term of the Rokhsar--Kivelson (RK) model \cite{rokhsar1988}, which turns each plaquette into a two level system diagonalized by the resonant states 
\begin{equation} 
|R_{\pm}\rangle=\frac{1}{\sqrt2}\left(|\vcenter{\hbox{\includegraphics[height=0.02\textheight]{figures/term1.pdf}}}\rangle \pm |\vcenter{\hbox{\includegraphics[height=0.02\textheight]{figures/term2.pdf}}}\rangle\right)
\label{eq:res_state}
\end{equation}
with energies $\mp J_p$ respectively. This term contributes at first order in perturbation theory, and therefore the strength of the microscopic parameter $J_p$ should be tuned to the emergent energy scales of the effective model if we want different terms in the Hamiltonian to compete. In the following, we will measure $J_p$ in units of the effective hopping $\kappa_{\mathrm{d}}=\kappa^2/2h$.

\begin{figure*}[t!]
    \centering
    \includegraphics[width=0.95\linewidth]{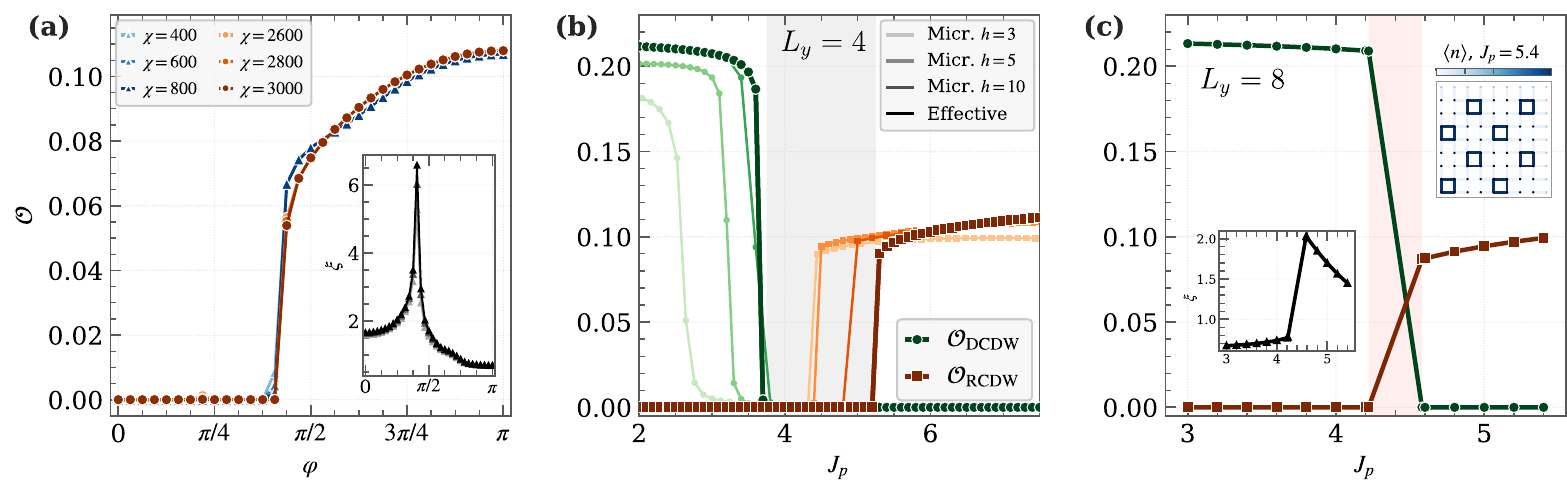}
    \caption{\textbf{(a)}: D-CDW order parameter as a function of the angle $\varphi$, for infinite cylinders of circumferences $L_y=4$ (blue) and $L_y=8$ (red). The results are consistent with a continuous transition at finite $\varphi$. The inset shows the iMPS correlation length, which peaks at the phase transition for $L_y=4$. \textbf{(b)}: local order parameters $\mathcal{O}_{\text{DCDW}}$ and $\mathcal{O}_{\text{RCDW}}$ as a function of $J_p$ at $\varphi=\pi$, for infinite cylinders of circumference $L_y=4$. The scan focuses on the critical region, revealing that the two ordered phases are interleaved by an intermediate region (gray shade). Results are compared to the full microscopic Hamiltonian \eqref{eq:2DFS}, showing that the same behavior qualitatively persists even at moderate values of $h$. \textbf{(c)}: same as panel \textbf{(b)}, for a cylinder of circumference $L_y=8$. For this system size the intermediate region (red shade) exhibits poor convergence of the iDMRG simulations at the maximum reached bond dimension $\chi=4000$, so that a refined study of the behavior of the order parameters is not possible. The converged numerical results outside of this region show clearly how the intermediate phase narrows down for larger systems. The insets show the correlation length across the transition (bottom-left) and the dimer density in the R-CDW phase (top-right).}
    \label{fig:phi_transition}
\end{figure*}

\section{Encoding of statistics}
\label{sec:statistics}
According to \cref{eq:qdh}, the statistics of the microscopic constituents is entirely encoded in the dimer hopping phase, which is a local quantity. To better understand how this works, in \cref{fig:phase_diagram}b we represent the original square lattice together with the ``dual'' $45^{\circ}$ rotated lattice obtained by connecting the centers of the links, with dimers on its vertices. The $\varphi$ phases acquired when hopping around a plaquette - or any other closed loop - always cancel out, so that the difference cannot be attributed to an emergent local magnetic flux. Further investigation reveals that the particular pattern of phases encodes the underlying statistics by keeping track of whether a dimer changes its orientation by $180^{\circ}$ after any sequence of hoppings that brings it back to its original position. An example of this is shown in \cref{fig:phase_diagram}c. Indeed, a change in the dimer orientation corresponds to swapping the positions of their microscopic constituents, and this must contribute a $\pi$ phase in the case of fermions. 

Given how simply $\varphi$ encodes the microscopic statistics, it is compelling to extend its range to arbitrary values between $0$ and $\pi$. The choice of how to do so is not unique, as for instance we could change the sign of $\varphi$ in one of the two terms in \cref{eq:qdh}, which makes no difference when $\varphi=0,\pi$ but does otherwise. The guiding principle that we adopt is the following: given that for $\varphi=0,\pi$ the net phase picked up when hopping around a closed loop is $0$, we require this to remain true for generic values of $\varphi$. In this way the only role of the phase is to encode a change in the orientation of a dimer, without contributing a magnetic flux. As argued above, this is exactly what the choice in \cref{eq:qdh} achieves. 

In terms of a microscopic model, this corresponds to each dimer being formed by a pair of particles with Abelian anyonic mutual statistics $\varphi$ \cite{wilczek1982}, but with bosonic statistics with respect to all the others\footnote{If all particles were of the same species, the dimers themselves would be Abelian anyons with statistical phase $2\varphi$, which is not captured by the effective model \cref{eq:qdh}.}. Such models are consistent, yet their lattice realization has not been studied in detail and it presents challenges, such as the introduction of auxiliary statistical Chern--Simons gauge fields \cite{ELIEZER199266, peng2026}. In the following we will treat \cref{eq:qdh} as a standalone dimer model, whose properties we wish to understand independently of its microscopic origin. It is important to stress that only the $\varphi=0,\pi$ limits have a clear interpretation in terms of the underlying $\Zt$ LGT.

\section{Ground states}
\label{sec:gs} 
The ground state of \eqref{eq:2DFS} in the large $h$ regime and $J_p=0$ differs drastically for bosons and fermions at half filling. In the former case confined dimers form a superfluid \cite{kebric2026}, while in the latter the repulsive interactions are sufficiently strong to stabilize a D-CDW state with the pattern shown in \cref{fig:phase_diagram}a, \ref{fig:phase_diagram}d and \ref{fig:phase_diagram}e \cite{borla2022}. In this section we first study how these two limits are connected, and then analyze the effect of a finite magnetic term. Numerical simulations are performed using the iDMRG algorithm on cylinders of size up to $L_y=8$, see \cref{app:numerical_methods} for details.

We perform a scan of the angle $\varphi$ and track the behavior of a local order parameter that detects the distinctive D-CDW density modulations. In the thermodynamic limit the pattern is 8-fold degenerate, with the different realizations being related to each other by translations in either direction and $\pi/2$ rotations. On the infinite cylinder geometry used in the numerical simulations the degeneracy is partially lifted and we find that a state with periodicity $4$ along the circumference and periodicity $2$ along the axis is energetically favored. This is captured by 
\begin{equation}
    \mathcal{O}_{\text{DCDW}}= \frac{1}{8} \sum_{\rr \,\in \mathcal{D} } \langle \hat n_{\rr, \hat y}^d\rangle \, e^{i \mathbf{Q}^{\mathrm{DCDW}}\cdot \mathbf{r}},
    \label{eq:op_mott}
\end{equation}
where the sum extends over a minimal $2\times4$ unit cell $\mathcal{D}$ which can accommodate the D-CDW pattern and $\mathbf{Q}^{\mathrm{DCDW}}=(\pi,\pi/2)$.
\cref{fig:phi_transition}a shows how the order parameter rises at an intermediate value $\varphi \approx 0.4 \pi$ both for $L_y=4$ and $L_y=8$. For the larger system size $\mathcal{O}_{\text{DCDW}}$ has a smoother behavior, compatible with a continuous phase transition. A divergence of the correlation length $\xi$ with respect to the bond dimension $\chi$ also occurs at the same value of $\varphi$, as shown in the inset for $L_y=4$. For $L_y=8$ the correlation length converges very slowly but at the largest achieved bond dimension $\chi=3000$ one can see a peak developing in the proximity of the quantum critical point, as discussed in more detail in \cref{app:numerical_methods}. At lower values of $\varphi$ the slow and monotonic convergence of $\xi$ with respect to $\chi$ indicates a gapless phase, consistent with our expectations. We note however that, due to the infinite cylinder geometry that we employ, the numerical results cannot fully characterize the nature of gapless phases, where the correlation length exceeds the circumference of the cylinder. In particular, we expect that the D-SF appears in this quasi-1d context under the guise of a gapless Luttinger liquid, which exhibits power-law correlations instead of true 2d off-diagonal long-range order. For the same reason, the 2d universality class of the quantum phase transition cannot be pinpointed with the methods at our disposal.

\begin{figure*}[t!]
    \centering
    \includegraphics[width=0.98\linewidth]{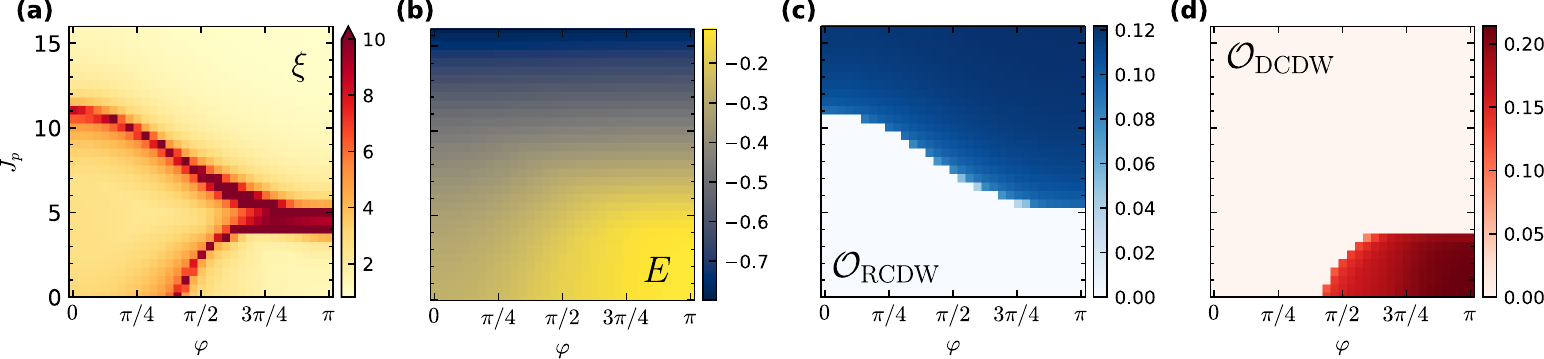}
    \caption{Numerical results for relevant ground state observables, in the $\varphi$-$J_p$ plane. \textbf{(a)}: peaks in the correlation length $\xi$ signal phase boundaries. \textbf{(b)}: ground state energy of the system. \textbf{(c-d)}: local order parameters \cref{eq:op_mott} and \cref{eq:op_plaq}, which detect the R-CDW and D-CDW ordered phases respectively. All simulations are run for an infinite cylinder of circumference $L_y=4$.}
    \label{fig:scan_J_theta}
\end{figure*}

We now investigate the role of a finite plaquette term. When $J_p \gg\kappa_\mathrm{d}$ the ground state is characterized by the formation of magnetically bound pairs. The $\hat R_\square$ term of \cref{eq:RKN} plays the role of a density operator for the resonating pairs and evaluates to $+1$ on fully resonating plaquettes. At half-filling, for instance, the formation of well separated resonant dimer pairs is reflected by an expectation value $\mathcal{R} = N_p^{-1}\sum_\square\langle \hat R_\square \rangle \approx 1/8$, with $N_p$ the total number of plaquettes. As already noticed in \cite{borla2022}, however, a large $J_p$ tends to favor a distinct state where dimers cluster together, corresponding to phase IV of the diagram in \cref{fig:phase_diagram}a. Our analysis focuses on the intermediate regime $U_\text{d} \approx J_p$, characterized by the interplay between the effective inter-dimer repulsion and magnetic binding. At $\varphi=\pi$, the D-CDW state is initially stable, and due to its gapped nature no pairing effect is visible at small finite $J_p$. At $J_p/\kappa_\text{d} \approx 5$ the system orders into a regular staggered pattern of resonating plaquettes at wave vector $\mathbf{Q}^{\mathrm{RCDW}}=(\pi/2,\pi/2)$, corresponding to phase III of \cref{fig:phase_diagram}a. This is captured by a local order parameter

\begin{equation}
  \mathcal{O}_{\mathrm{RCDW}}
  \;=\;
  \frac{1}{16}
  \,
    \sum_{\rr \,\in \mathcal{D} }
    \langle \hat R_\square\rangle_{\rr}\,
    e^{\,i\,\mathbf{Q}^{\mathrm{RCDW}}\cdot\rr}
  \,
  \label{eq:op_plaq}
\end{equation}
where the sum extends over a $4\times4$ unit cell. The interpretation is the following: while a large $J_p$ overcomes the repulsion between dimers and allows them to pair up, the remnant inter-pair repulsive interactions are sufficiently strong to stabilize a gapped insulating state. This can be seen as a charge density wave of resonating pairs (R-CDW), which are the emergent fundamental constituents in this regime. The D-CDW and R-CDW states are an example of two gapped ordered phases which realize different translational symmetry breaking patterns. How the two connect to each other as $J_p$ is increased is a non-trivial question to answer, which can lead to a number of distinct possibilities. Our numerical simulations on infinite-cylinders of circumference $L_y=4$, shown in \cref{fig:phi_transition}b, suggest that an intermediate region characterized by vanishing order parameters, a large iMPS correlation length and uniform dimer density, appears. As the circumference size is increased to $L_y=8$ the extent of this region reduces, as shown in \cref{fig:phi_transition}c. For this system size numerical simulations in the intermediate region exhibit very poor convergence even at a large bond dimension $\chi=4000$, which we attribute to the presence of near-degenerate ground states with competing order. Because of this it is not possible to determine more precisely the near-critical behavior of the order parameters, the exact boundaries of the intermediate phase and whether -unlike the $L_y=4$ case- there is a coexistence region. Since the narrowing of the intermediate phase appears in any case robust, we can speculate that this could shrink to a point in the thermodynamic limit, which would generically correspond to a direct first order transition between the ordered phases. A more exotic possibility is the appearance a deconfined quantum critical point, characterized by emergent fractionalized degrees of freedom which become confined and therefore ``invisible'' on either side of the transition \cite{senthil2004, senthil2004review, wang2017deconfined, senthil2023deconfinedquantumcriticalpoints}. In \cref{fig:phi_transition}b we also compare the effective model and the microscopic model \cref{eq:2DFS} at increasing values of $h$ within the confined phase, showing that a similar qualitative picture persists at moderate values of the electric coupling.

\cref{fig:scan_J_theta} shows a full iDMRG scan of the $\varphi$-$J_p$ plane for an infinite cylinder of circumference $L_y=4$, revealing the connectivity of the phase diagram. Phase boundaries are signaled by peaks in the iMPS correlation length $\xi$. Ordered phases are distinguished through the local order parameters $\mathcal{O}_{\text{DCDW}}$ and $\mathcal{O}_{\text{RCDW}}$. The transition shown in \cref{fig:phi_transition}a appears to survive at finite magnetic coupling, with the critical point drifting towards larger values of $\varphi$ as $J_p$ is increased. The narrow intermediate phase that occurs between the two ordered states extends to values $\varphi < \pi$ and, interestingly, is smoothly connected to the D-SF. The plaquette order occupies the whole large $J_p$ region up to a value $J_p^{\text{clust.}}/\kappa_\text{d} \approx 18$, where the system undergoes a first order transition to a clustered phase (not displayed in \cref{fig:scan_J_theta}). The magnetic term washes out the differences between dimers with bosonic and fermionic constituents. We notice however that the R-CDW ordering transition occurs at a significantly higher value $J_p^{\text{crit}}|_{\varphi=0} \approx 2J_p^{\text{crit}}|_{\varphi=\pi}$, consistent with the idea that dimers with bosonic constituents have larger kinetic energy compared to their fermionic counterparts. This argument can be made more precise by looking at the single-dimer band structure resulting from \cref{eq:qdh}, discussed in \cref{app:single_dimer}. Solving the two-dimer problem, on the other hand, correctly predicts the formation of magnetic bound states for values of the coupling which qualitatively agree with the many-body results shown in \cref{fig:scan_J_theta}, see \cref{app:two_dimers} for details.

\section{Dynamical signatures}
\label{sec:dyn}
\begin{figure*}[t!]
        \centering
        \includegraphics[width=0.95\linewidth]{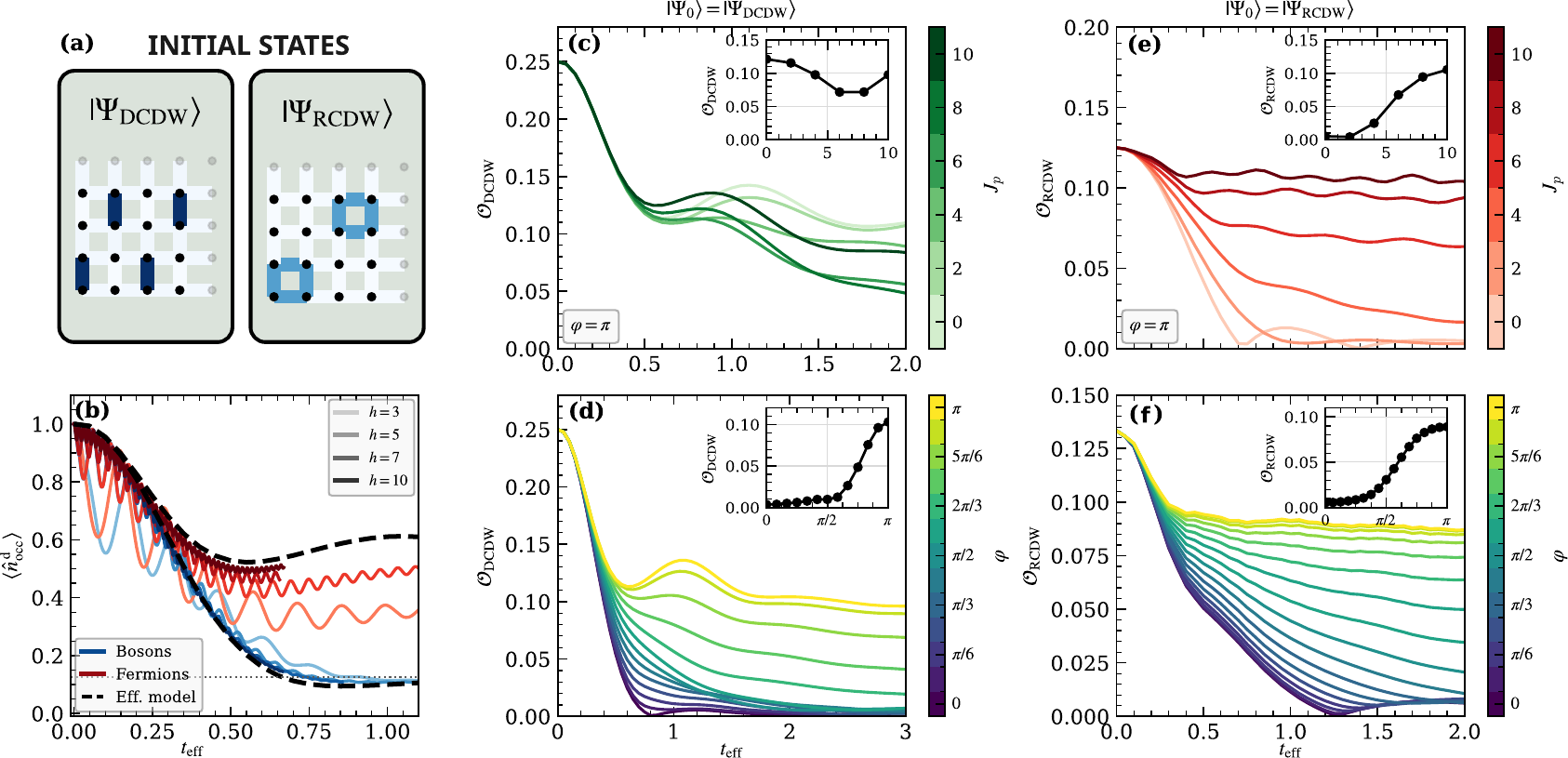}
        \caption{Dynamical signatures of quantum order. \textbf{(a)}: initial configurations $|\Psi^0_{\mathrm{DCDW}} \rangle$ and $|\Psi^0_{\mathrm{RCDW}} \rangle$ used in the quench protocol. \textbf{(b)}: comparison of quenches of $|\Psi^0_{\mathrm{DCDW}} \rangle$ in the microscopic model \eqref{eq:2DFS} with bosonic (blue) and fermionic (red) matter deep in the confined phase, and in the effective model at $\varphi=0$ and $\pi$. At large $h$, the effective model description becomes exact. The initial order, measured by the matter density on an initially occupied link, survives at long times only for fermionic dimers. \textbf{(c)}: quenches of $|\Psi^0_{\mathrm{DCDW}} \rangle$ for different values of $J_p$ at $\varphi=\pi$. A finite magnetic term partially melts the initial order over the accessible timescales. The decay rate of the order parameter is non-monotonic in $J_p$. \textbf{(d)}: quenches of $|\Psi^0_{\mathrm{DCDW}} \rangle$ for different values of the statistical angle $\varphi$. The inset shows how a qualitative change in the dynamics occurs at $\varphi_c \approx 0.58 \pi$, after which the time averaged D-CDW order parameter attains a finite value. \textbf{(e)}: quenches of $|\Psi^0_{\mathrm{RCDW}} \rangle$ for different values of $J_p$ at $\varphi=\pi$. The R-CDW order melts as $J_p$ is decreased. Compared to the quenches in panel (c), the decay of the order parameter is sharper, faster and monotonic. \textbf{(f)}: quenches of $|\Psi^0_{\mathrm{RCDW}} \rangle$ for different values of the statistical angle $\varphi$, at fixed $J_p = 7.5$. Here the dynamical order parameter starts to grow at $\varphi_c \approx 0.5 \pi$.}
    \label{fig:dynamics}
\end{figure*}

Having established the existence of dimer phases with distinctive crystalline order, in this section we propose a quench protocol to detect their signatures dynamically, and we test it with tensor network simulations. We consider simple initial states that realize the two ordered patterns shown in \cref{fig:dynamics}a, and quench them to different parameter regimes. For the D-CDW order, we define the corresponding initial product state $|\Psi^0_{\mathrm{DCDW}} \rangle$ by placing dimers on the appropriate links. For the plaquette order, we use the initial state $|\Psi^0_{\mathrm{RCDW}} \rangle$ prepared in the following way: we start from a product state with pairs of dimers $|vv\rangle$ on parallel neighboring vertical links, arranged to reproduce the desired pattern. Then on every occupied plaquette we apply a rotation $R_x(-\pi/2)$ in the $hh$-$vv$ subspace, giving $(|vv\rangle+i|hh\rangle)/\sqrt{2}$. Finally, a local diagonal phase gate $e^{-i \frac{\pi}{2} \hat{n}}$ on one horizontal bond of each occupied plaquette is applied to produce the correct superposition. The initial state is then evolved with the effective Hamiltonian or when possible, for comparison, with the microscopic Hamiltonian \cref{eq:2DFS}. For the numerical simulations we employ the infinite TDVP algorithm for the microscopic model and the finite TDVP algorithm with $L_x=16$ for the effective model, both on cylinders of circumference $L_y=4$, see \cref{app:numerical_methods} for details.

As a first application of our protocol, we test the difference between dimers with bosonic and fermionic constituents. To do so we set $J_p=0$, and choose $|\Psi^0_{\mathrm{DCDW}} \rangle$ as the initial state. This has a sizable overlap with the ground state of the fermionic Hamiltonian, and is therefore expected to show near-equilibrium behavior and persistent ordering over intermediate time scales. This is to be contrasted with the bosonic case, where such initial state is far from equilibrium and should exhibit fast dynamics, relaxing to a final state where the translational symmetry is restored.  In \cref{fig:dynamics}b the evolution under the effective model is compared to the full dynamics governed by the microscopic Hamiltonian \eqref{eq:2DFS} at different values of $h$. We track the expectation value of $\hat n^d$ on one of the initially occupied links. As the electric coupling is increased and the times are rescaled by the effective energy scale $\kappa^2/2h$, the curves for both bosons and fermions collapse towards the corresponding effective model ($h=\infty$) prediction. In the former case, the dimer density decays to $\langle \hat n_d \rangle=1/8$, which is the value expected in a translationally invariant state at half-filling of microscopic constituents. For fermionic matter, on the other hand, the particle density on the initially occupied link maintains a large value up to the maximum timescale we were able to probe, a signature of persistent ordering. 

Next, we test the stability of the ordered phases at $\varphi=\pi$ after quenching to finite values of the magnetic term $J_p$. We consider first $|\Psi^0_{\mathrm{DCDW}} \rangle$: as shown in \cref{fig:dynamics}c, the order melts in a non-monotonic fashion, with the largest decay rate of $\mathcal{O}_{\text{DCDW}}$ attained at the intermediate values $J_p/\kappa_\text{d}=6$-$8$. Interestingly, this non-monotonic behavior tracks the three ground state regimes of \cref{fig:phi_transition}b. We note that due to the entangling nature of the plaquette term, numerical simulations at finite $J_p$ are challenging and even within the effective model we are only able to reach relatively short times, at which no complete drop in the order parameter is observed. Direct simulations of the microscopic model over the same rescaled times are prohibitive, so that a comparison with the full dynamics at lower $h$ is only possible up until $t_{\mathrm{eff}}\approx 1$, where differences are not visible. 
Similarly, we test the stability of the plaquette order of the initial state $|\Psi^0_{\mathrm{RCDW}} \rangle$ by decreasing $J_p$. The results in \cref{fig:dynamics}e show that the order parameter exhibits a progressively quicker decay at lower $J_p$, and no intermediate regime is visible. 

Following up on the idea that the angle $\varphi$ in \cref{eq:qdh} interpolates between the two types of dimers, we investigate how the quench dynamics of ordered states depends on it. We consider first $|\Psi^0_{\mathrm{DCDW}} \rangle$ at $J_p=0$, as shown in \cref{fig:dynamics}d. Similarly to how an equilibrium quantum critical point is observed at finite $\varphi$ in \cref{fig:phi_transition}a, we identify a dynamical value of $\varphi$ at which the long time behavior of the system changes qualitatively from an ordered symmetry broken density pattern to a homogeneous one. In the inset we show how the time-averaged staggered order parameter 
\begin{equation}
\mathcal{O}_{\text{DCDW}}^{\text{dyn}}=\frac{1}{t_{\text{max}}-t_0}\int_{t_0}^{t_{\text{max}}} \mathcal{O}_{\text{DCDW}}(t) \, dt,
\end{equation}
where $t_0$ is chosen to exclude the initial transient, exhibits a behavior reminiscent of a conventional order parameter, rising at a comparatively larger value $\varphi_c\approx 0.58 \pi$. 
To probe the stability of $|\Psi^0_{\mathrm{RCDW}} \rangle$ we set instead $J_p/\kappa_\text{d}=7.5$, corresponding to an ordered ground state at $\varphi=\pi$ but not at $\varphi=0$. The order indeed persists for values of $\varphi$ close to $\pi$, and the corresponding time-averaged order parameter 
\begin{equation}
\mathcal{O}_{\text{RCDW}}^{\text{dyn}}=\frac{1}{t_{\text{max}}-t_0}\int_{t_0}^{t_{\text{max}}} \mathcal{O}_{\text{RCDW}}(t) \, dt
\end{equation}
exhibits a smooth rise near $\varphi=0.5 \pi$. 

Our protocol demonstrates how signatures of equilibrium ordered phases can be inferred dynamically by looking at the behavior of appropriate order parameters over short, accessible timescales. Specializing to $\varphi=0,\pi$, the results show the impact that the microscopic statistics (bosonic vs fermionic) has on the time evolution of otherwise indistinguishable and identically interacting dimer configurations.

\section{Conclusions and outlook}
\label{sec:concl}
In this article we studied how the statistics of microscopic constituents can affect the ground state and dynamical properties of dimer models describing the confined regime of certain lattice gauge theories. Building on the effective dimer model formulation of a strongly coupled $2+1$D $\Zt$
LGT, we find that this statistics enters exclusively through a local hopping phase $\varphi$, while the interactions are unaffected. A non-local property is therefore reduced to a local parameter of the emergent low-energy theory, while retaining striking physical effects. Although the particular mechanism by which this happens is model-specific, the underlying idea is that confinement limits the possibilities for particle exchange. This is generalizable to composite objects that are more complex than unit length dimers, and to any other gauge theory where electrically bound mesons are well defined. Taking $\varphi$ to be a continuous variable between $0$ and $\pi$ has a natural interpretation in terms of the exchange statistics of the two particles composing individual dimers, providing an example of ``interpolation'' between bosons and fermions. 

From the point of view of ground state physics, the interplay between a $\varphi$ dependent hopping, emergent repulsive interactions and a RK kinetic term produces a quantum phase diagram which includes different crystalline states. In particular, we find an exotic gapped phase characterized by $(\pi/2,\pi/2)$ spatial ordering, formed by magnetically bound dimer pairs locked into a crystalline pattern by the inter-dimer repulsion. The novelty of this dimer phase is two-fold. Firstly, from the perspective of the microscopic $\Zt$ LGT, we witness a two-tier hierarchy of emergent degrees of freedom, tied to distinct fundamental microscopic processes. The electric field confines $\Zt$ charges into unit-length dimers, which are in turn bound into resonating pairs by the magnetic term. Secondly, this state provides a natural example of how crystalline order can occur in dimer models at finite filling as a consequence of effective repulsive interactions. The interactions pictured in \cref{fig:phase_diagram}d, while intricate, are natural for dimers composed by two elementary particles on the square lattice. They arise exclusively from virtual processes involving the constituents, and are therefore easily generalizable.

Another point of interest is the precise nature of the quantum critical points occurring in the phase diagram \cref{fig:phase_diagram}a. This cannot be pinned down with the methods employed in this article, as the system size limitations inherent to MPS in 2d do not allow a proper finite size scaling analysis. In the vicinity of the $\varphi=\pi$ axis, the D-CDW and R-CDW phases are interleaved by a small gapless region, smoothly connected to the D-SF. Our data indicates that as the system size is increased this intermediate region narrows down, potentially collapsing onto a single line in the thermodynamic limit. Generically, a first order transition between the two competing orders is expected. Other possibilities include a deconfined quantum criticality separating phases which break distinct lattice symmetries. A detailed study of this transition requires fully 2d numerical methods. While to the best of our knowledge Monte Carlo techniques are affected by a sign problem when $\varphi\neq0$, promising alternative approaches include finite and infinite projected entangled pair states (PEPS) \cite{Verstraete01032008, naumann2024peps} and neural quantum states (NQS) \cite{Carleo_2017, Lange_2024, kufel2025}.

Motivated by the increased availability and efficiency of quantum simulation platforms, we provided a simple dynamical protocol to probe the ordered phases of the model. Persistence or melting of initial ordered patterns can be put in qualitative correspondence with ground state phases realized by the Hamiltonian. As a consequence, we find that identical dimer configurations can retain or lose their order depending only on the statistics of their microscopic constituents. As the initial states are easily prepared and the fermionic statistics enters solely through a local phase, the protocol can be implemented in near-term quantum simulators, giving access to strongly entangled regimes where tensor-network methods become less effective.

\begin{acknowledgments}
We thank Sergej Moroz and Jesse J.~Osborne for fruitful discussions. U.B.~and J.C.H.~acknowledge funding by the Max Planck Society, the Deutsche Forschungsgemeinschaft (DFG, German Research Foundation) under Germany’s Excellence Strategy – EXC-2111 – 390814868, and the European Research Council (ERC) under the European Union’s Horizon Europe research and innovation program (Grant Agreement No.~101165667)—ERC Starting Grant QuSiGauge. This work is part of the Quantum Computing for High-Energy Physics (QC4HEP) working group. J.C.H.~acknowledges the hospitality of the Aspen Center for Physics while this paper was being completed.
\end{acknowledgments}

\appendix
\tightmathspacing
\crefalias{section}{appendix}

\crefname{section}{Appendix}{Appendices}
\Crefname{section}{Appendix}{Appendices}
\crefname{subsection}{Appendix}{Appendices}
\Crefname{subsection}{Appendix}{Appendices}
\section{Single dimer band structure}
\label{app:single_dimer}

\begin{figure}
    \centering
    \includegraphics[width=\linewidth]{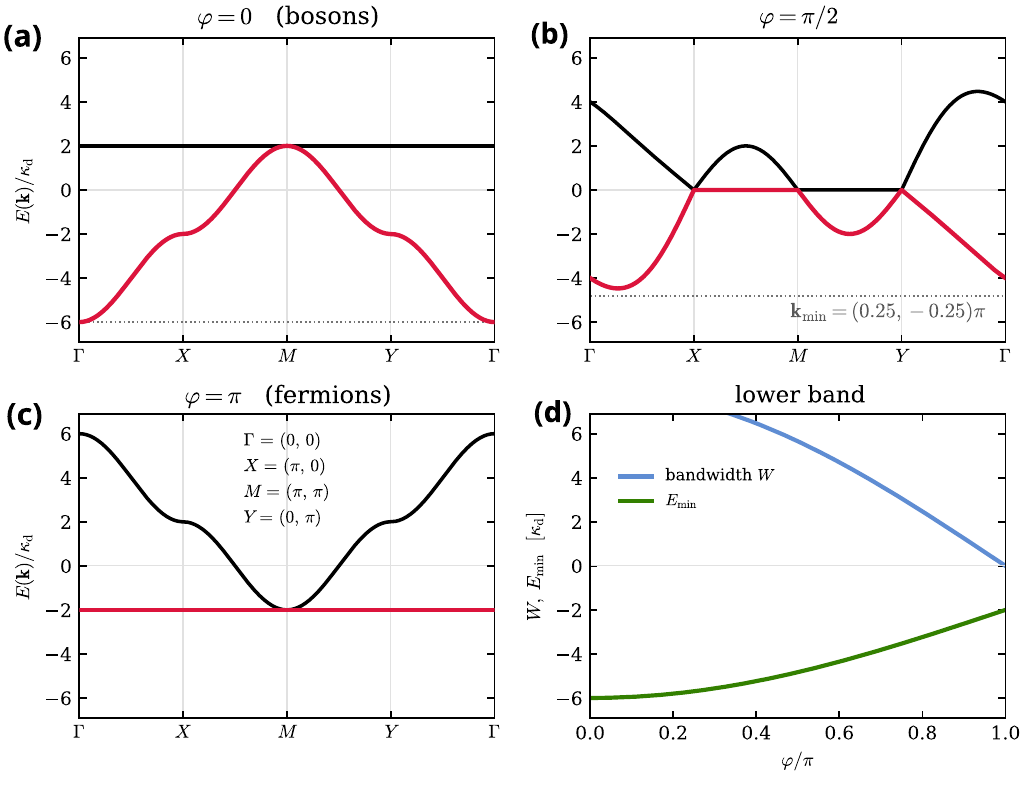}
    \caption{\textbf{(a-c)}: the two bands of the hopping Hamiltonian \eqref{eq:qdh}, drawn across $\mathbf{k}$ cuts connecting the high symmetry points of the Brillouin zone, for different values of $\varphi$. The extremal values $\varphi=0$ and $\varphi=\pi$ both feature a flat band, at the top and bottom of the spectrum respectively. \textbf{(d)}: band minimum and width of the lowest band in the single-dimer dispersion.}
    \label{fig:bands}
\end{figure}

From the hopping Hamiltonian \cref{eq:qdh} it is possible to derive the single dimer band structure. Since dimers live on a dual lattice with twice as many sites as the original one, the Bloch Hamiltonian is a $2 \times 2$ matrix in momentum space, leading to two distinct energy bands. Cross sections are shown in \cref{fig:bands}a-c, together with the minimum and width of the lower band in \cref{fig:bands}d. As noted already in \cite{borla2022}, at $\varphi=\pi$ the lowest band is completely flat and touches the upper dispersive band at $k_x=k_y=\pi$. For $\varphi=0$ we also observe a flat band, but this time at the top of the spectrum. Even in the absence of interactions the hardcore constraints make the problem non-free, so that many-body properties cannot generally be derived from the single particle dispersion unless the system is sufficiently dilute. Nonetheless, the markedly different band structure near the energy minima allows one to draw at least some qualitative conclusions about the many-particle ground state. At $\varphi=0$ the lowest band is dispersive and has a single minimum at $\mathbf{k}=0$. The energy $E_\text{min}=-6\kappa_d$ and bandwidth $\approx 8\kappa_d$ are relatively large compared to the typical interaction strength $\approx 2\kappa_d$, and in this situation hardcore bosons are generically expected to condense, consistent with the observed D-SF phase. At $\varphi=\pi$, on the other hand, the vanishing bandwidth makes the system sensitive to small interactions, which stabilize the crystalline pattern shown in the main text. 

\section{Two dimer problem and bound states}
\label{app:two_dimers}

\begin{figure}
    \centering
    \includegraphics[width=\linewidth]{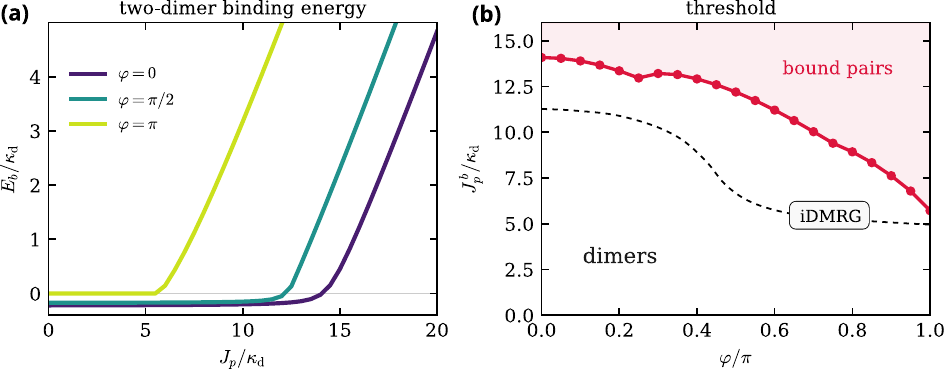}
    \caption{\textbf{(a)}: binding energy of the two dimer problem, computed with exact diagonalization on lattices of size up to $10 \times 10$. \textbf{(b)}: pairing threshold, defined as the value of $J_p$ for which the bound state has a lower energy compared to two independent dimers.}
    \label{fig:bound_states}
\end{figure}

The two-dimer problem can be solved with exact diagonalization on lattices large enough that the results are not affected by finite size limitations. We work directly in the constrained Hilbert
space, formed by dimer configurations which are allowed by the extended hardcore constraint, and diagonalize the
full effective Hamiltonian $H_{\text{hop}}+H_{\text{int}}+H_{\text{plaq}}$ on an $L\times L$ torus. The basis contains $2L^2(2L^2-7)/2$ states, i.e. $7\,744$ states for $L=8$ and $19\,300$ for $L=10$. With this method it is possible to study whether two dimers form a bound state at finite $J_p$. This happens if the two-body ground state
lies below the threshold for two independent dimers, i.e. if
\begin{equation}
  E_b \;=\; 2E_{\min}(\varphi)\;-\;E_2(\varphi,J_p)\;>\;0,
  \label{eq:binding}
\end{equation}
with $E_{\min}$ the single-dimer band minimum discussed in \cref{app:single_dimer}. \cref{fig:bound_states}a shows that the binding energy is flat and non-positive below a sharp threshold $J_p^{\,b}(\varphi)$, and grows linearly
above it. \cref{fig:bound_states}b gives the threshold
across the full range of $\varphi$, compared to the iDMRG curve corresponding to a transition to the R-CDW phase. Inspection of the bound state confirms its nature: at $\varphi=\pi$ and
$J_p=8\,\kappa_{\text{d}}$ we find $\langle\sum_\square \hat R_\square\rangle=0.78$. The magnetically bound pair detected numerically through many-body simulations in the main text has therefore a strict two-body bound state counterpart. Such bound states are expected to be heavy, i.e. to have small kinetic energy compared to the interaction scale. This facilitates the formation of a crystalline pattern at commensurate fillings.

\section{Tensor network methods and convergence}
\label{app:numerical_methods}

\begin{figure}
    \centering
    \includegraphics[width=\linewidth]{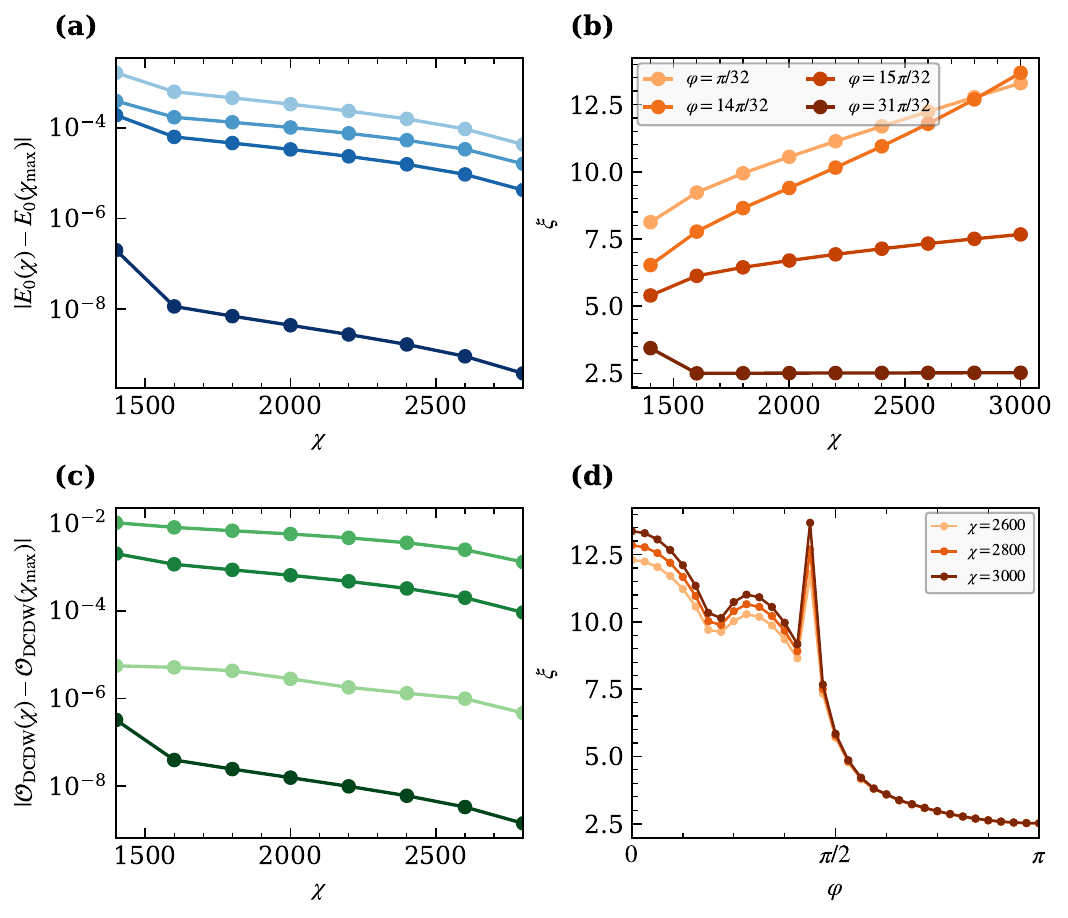}
    \caption{Convergence tests for the $L_y=8$ simulations shown in \cref{fig:phi_transition}a. Panels \textbf{(a)} and \textbf{(c)} show the convergence of the ground state energies and order parameter $\mathcal{O}_{\mathrm{DCDW}}$ by comparing intermediate values of the bond dimension to the maximum $\chi=3000$.
    Panel \textbf{(b)} shows the behavior of the correlation length (plotted in \textbf{(d)} as a function of $\varphi$) as a function of the bond dimension for different $\varphi$ corresponding to the gapless D-SF, near-critical and ordered regions.}
    \label{fig:idmrg_conv}
\end{figure}

Our numerical work relies on a combination of different tensor-network based techniques. The effective model is implemented in \texttt{TeNPy} \cite{tenpy2024}, which implements the infinite density matrix renormalization group (iDMRG) algorithm for ground state optimization and the finite time dependent variational principle (TDVP) for time evolution. The microscopic $\Zt$ LGT model \cref{eq:2DFS} on the other hand is always simulated using \texttt{mptoolkit} \cite{mptoolkit}, which also implements the infinite version of the TDVP algorithm. Both for the infinite and finite case we use periodic boundary conditions in the $\hat y$ direction, corresponding to a cylindrical geometry.

\subsection*{Convergence of ground state simulations}
The most numerically demanding ground state result shown in the main text is the $L_y=8$ simulation shown in \cref{fig:phi_transition}a, near the critical point. In \cref{fig:idmrg_conv} we show convergence of the energy, correlation length and order parameter with the bond dimension $\chi$ at values of $\varphi$ near and far from the critical point. At the maximum achieved bond dimension $\chi=3000$, convergence of local observables is satisfactory, while the iMPS correlation length still exhibits noisy behavior which does not allow one to distinguish the different regimes clearly. In particular, while the smooth increase at small $\varphi$ is expected in the gapless D-SF phase, we see an unexpected non-monotonic behavior to the left of the critical point, which is not captured by any of the other observables. Deep in the gapped phase at large $\varphi$, $\xi$ converges quickly to a finite value.

\begin{figure}
    \centering
    \includegraphics[width=\linewidth]{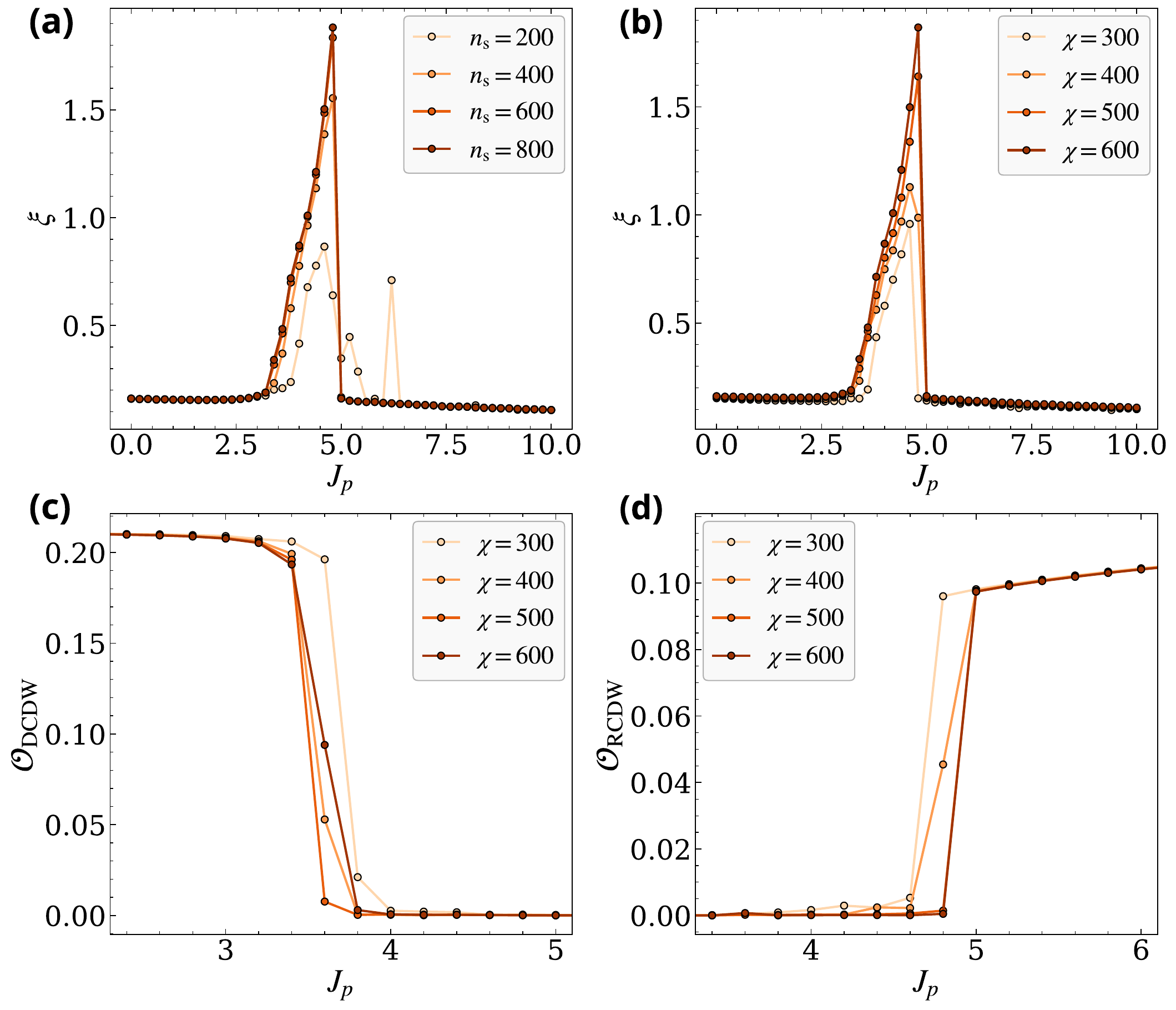}
    \caption{Convergence tests for the equilibrium properties of the microscopic model in Eq.~\eqref{eq:2DFS} with $h=10$. \textbf{(a)}: convergence of the correlation length $\xi$ for increasing number of iDMRG sweeps $n_s$, at fixed bond dimension $\chi=600$. \textbf{(b)}: convergence of the correlation length $\xi$ as the bond dimension $\chi$ is increased. \textbf{(c)-(d)}: convergence of different observables in bond dimension $\chi$.}
    \label{fig:conv_test}
\end{figure}

In the microscopic model at $L_y=4$, we study in detail the convergence of the correlation length to pinpoint the location of the intermediate gapless phase in \cref{fig:phi_transition}b. We ensure convergence of iDMRG by monitoring $\xi$ as a function of the number of sweeps. More specifically, we choose as initial state the product state $\ket{\Psi^0_{\mathrm{DCDW}}}$, and fix a maximum value of the bond dimension $\chi$. We then carry out $60$ iDMRG sweeps with the microscopic Hamiltonian in Eq.~\eqref{eq:2DFS} with mixing factor set to $10^{-5}$, followed by another $60$ sweeps with mixing factor $10^{-7}$. Then, we perform a total of $n_s=800$ sweeps with no mixing factor, saving the correlation length after every $20$ sweeps. The results of this procedure for $\chi=600$ and $h=10$ are shown in Fig.~\ref{fig:conv_test}(a), where we plot the correlation length $\xi$ as a function of the magnetic coupling $J_p$, for increasing number of sweeps $n_s$. Convergence in bond dimension is instead shown in Fig.~\ref{fig:conv_test}(b)-(d), where we plot the correlation length $\xi$ and the two order parameters $\mathcal{O}_{\text{DCDW}}$ and $\mathcal{O}_{\text{RCDW}}$ for increasing bond dimension $\chi$. Panel (b) shows the correlation length increasing with $\chi$ in the region $J_p\in(3,5)$, which is consistent with the gapless phase observed in the effective model.

\subsection*{Convergence of dynamics simulations}

\begin{figure}
    \centering
    \includegraphics[width=\linewidth]{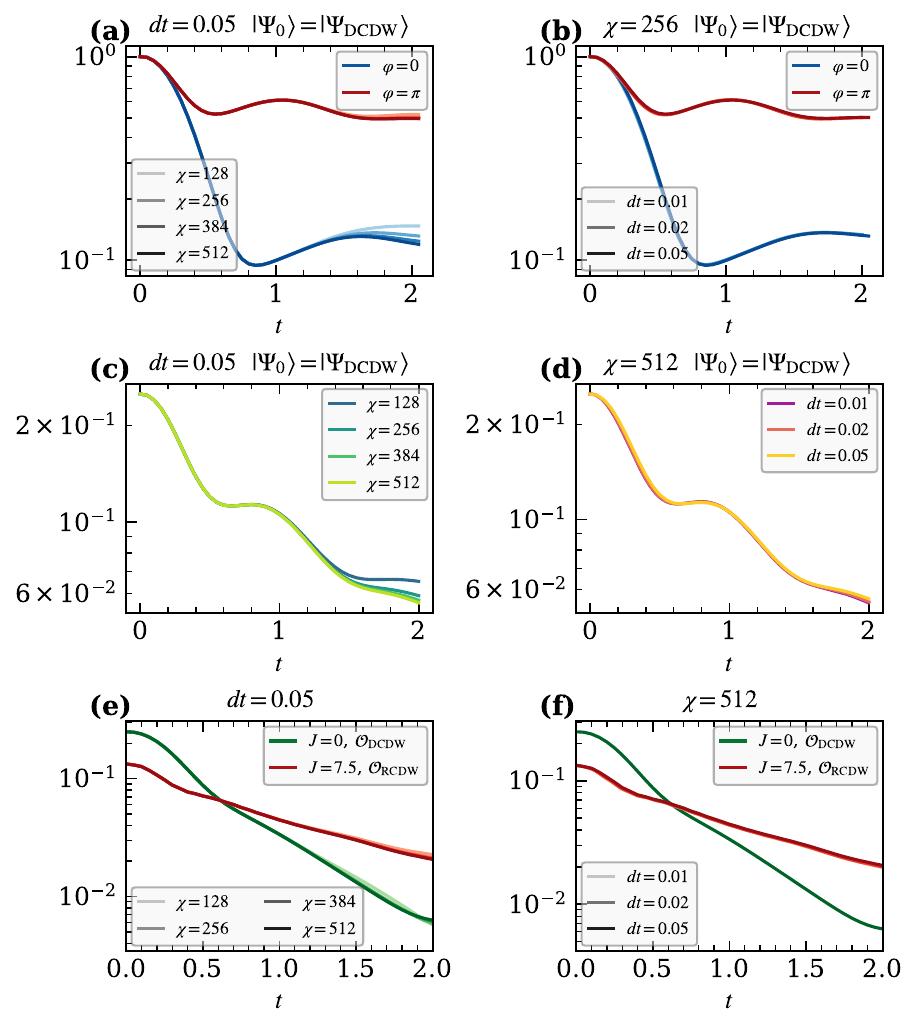}
    \caption{Convergence plots for the finite TDVP simulations of the effective model dynamics used in \cref{fig:dynamics}. \textbf{(a,b)}: quenches of $|\Psi_{\mathrm{DCDW}}\rangle$ to $\varphi=0,\pi$ at $J_p=0$, keeping $dt$ fixed and increasing $\chi$ ad vice-versa. \textbf{(c,d)}: quenches of $|\Psi_{\mathrm{DCDW}}\rangle$ to $J_p=6$ at $\varphi=\pi$. \textbf{(e,f)}: quenches of $|\Psi_{\mathrm{DCDW}}\rangle$ (green) and $|\Psi_{\mathrm{RCDW}}\rangle$ (red) to $\varphi=\pi/2$, for $J_p=0, 7.5$ respectively.}
    \label{fig:tdvp_conv}
\end{figure}

The time-evolution simulations under the effective model, shown in \cref{fig:dynamics}, are performed using the TDVP algorithm on finite cylinders of $L_x=16$ and $L_y=4$. We check the convergence of these simulations with respect to the MPS bond dimension $\chi$ and timestep $dt$. Results of the convergence tests for selected quenches relevant to the main text figures are shown in \cref{fig:tdvp_conv}.

Regarding the iTDVP algorithm used for the microscopic model, we run our simulations with a fixed timestep $dt$ and a fixed maximum bond dimension $\chi$, and check convergence in both parameters for the effective time intervals we are interested in. 
\begin{figure}[ht!]
    \centering
    \includegraphics[width=\linewidth]{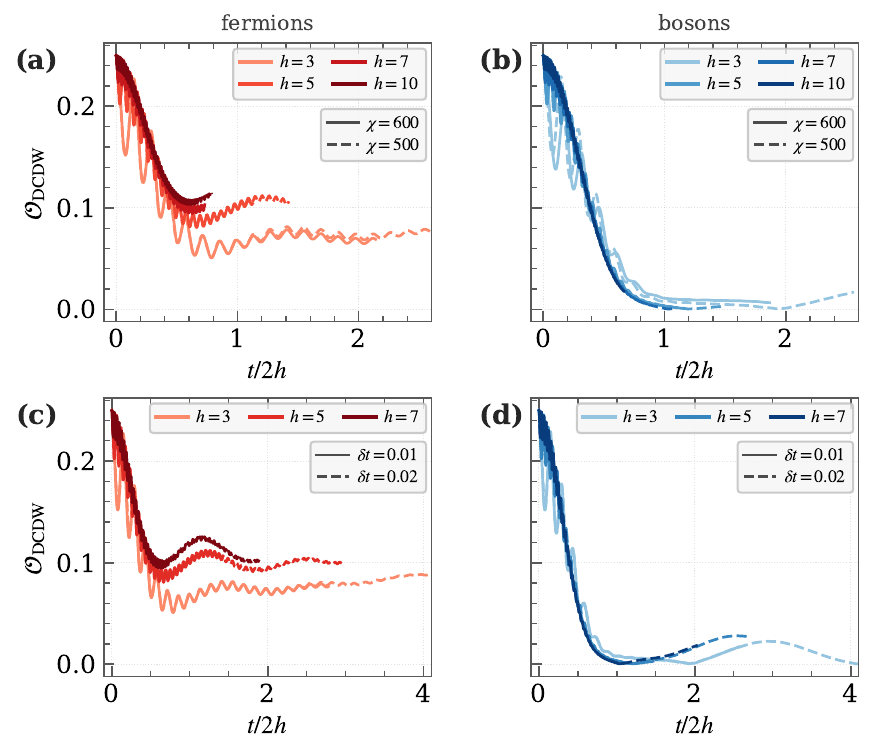}
    \caption{Convergence tests for the quench protocol described in Sec.~\ref{sec:dyn} with $J_p=0$, for both fermionic (red curves) and bosonic matter (blue curves) and different values of $h$. \textbf{(a)-(b)}: convergence in bond dimension $\chi$ for fixed time step $dt=0.01$. \textbf{(c)-(d)}: convergence in time step for fixed bond dimension $\chi=500$.}
    \label{fig:dynamics_convergence_J0}
\end{figure}
Fig.~\ref{fig:dynamics_convergence_J0} shows the order parameter $\mathcal{O}_{\text{DCDW}}$ as a function of rescaled time $t_{\mathrm{eff}} = t/2h$ for both fermionic (red curves) and bosonic matter (blue curves), for different values of $h$, with the magnetic coupling $J_p=0$. Panels (a) and (b) display convergence with bond dimension $\chi$ at fixed $dt=0.01$, while panels (c), (d) show convergence with time step $dt$ at fixed $\chi=500$. Note that in these last panels the curve for $h=10$ is missing, since a $dt=0.02$ causes the simulations to break almost immediately. The reason is that the (rescaled) frequency of the small oscillations of the order parameter scales linearly with $h$, meaning the absolute time step $dt$ should scale as $O(h^{-2})$ to capture them correctly. This makes the higher-$h$ simulations the hardest to compute with the microscopic model. We also note that convergence in $\chi$ is stronger for fermions, which is expected since in this case the chosen initial state has a large overlap with the ground state of the Hamiltonian, thus requiring a smaller bond dimension to accurately represent its dynamics compared to bosons.

\section{Validity of the dimer approximation}
\label{app:dimer_approx}
\begin{figure}[t!]
    \centering
    \includegraphics[width=0.98\linewidth]{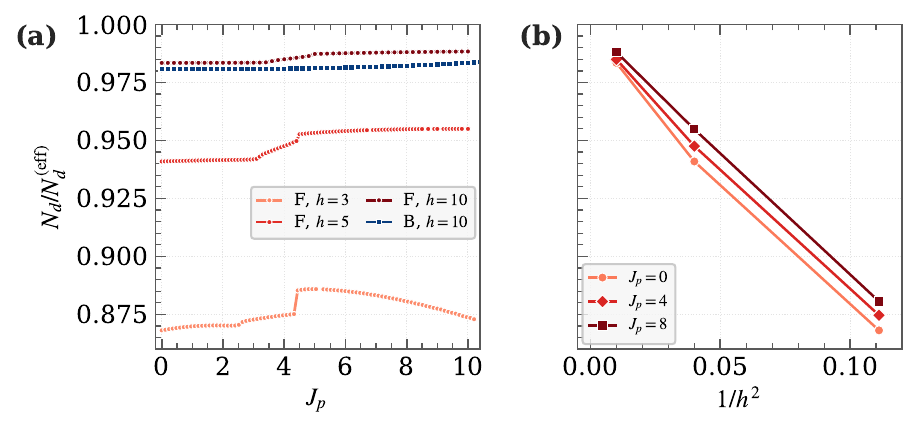}
    \caption{\textbf{(a)}: Validity of the unit-length dimer approximation in the microscopic model as $h$ increases, measured by the ratio between the total number of dimers in the microscopic model $N_d$ and its $h\to\infty$ value $N_d^{\mathrm{(eff)}}$, as a function of $J_p$. \textbf{(b)}: For fixed $J_p$, the ratio approaches unity linearly in $1/h^2$, which is the emergent perturbative energy scale.}
    \label{fig:dimer_index}
\end{figure}
We now show how well the dimer model approximates the underlying microscopic model as $h$ increases. To this end, we compute the number of length one dimers in the microscopic model as
\begin{equation}
    N_d =\sum_{\mathbf{r}, \eta} \langle \hat n_{\rr, \eta}^d\rangle, \qquad \hat n_{\rr, \eta}^d = \hat n_{\rr}\frac{1-\hat\sigma^x_{\rr,\eta}}{2}\hat n_{\rr+\eta}
\end{equation}
where the sum is over a unit cell for the infinite cylinder geometry and over the whole system for finite systems. In the effective model the dimer number is strictly fixed by the total microscopic particle number: $N_d=N/2$. This is not the case in the microscopic model, as a given number of constituents can potentially combine into longer dimers. In \cref{fig:dimer_index}a we plot the ratio between this quantity and its $h\to\infty$ value $N_d^{\mathrm{(eff)}}$ as a function of $J_p$ and for increasing values of $h$. The results are consistent with Fig. \ref{fig:dynamics}(b), where we observe a good quality approximation already at $h=10$. \cref{fig:dimer_index}b shows how the deviations from the perfect dimer approximation $N_d/N_d^{\text{eff}}=1$ are linear in the emergent energy scale $1/h^2$.


\begin{thebibliography}{73}%
\makeatletter
\providecommand \@ifxundefined [1]{%
 \@ifx{#1\undefined}
}%
\providecommand \@ifnum [1]{%
 \ifnum #1\expandafter \@firstoftwo
 \else \expandafter \@secondoftwo
 \fi
}%
\providecommand \@ifx [1]{%
 \ifx #1\expandafter \@firstoftwo
 \else \expandafter \@secondoftwo
 \fi
}%
\providecommand \natexlab [1]{#1}%
\providecommand \enquote  [1]{``#1''}%
\providecommand \bibnamefont  [1]{#1}%
\providecommand \bibfnamefont [1]{#1}%
\providecommand \citenamefont [1]{#1}%
\providecommand \href@noop [0]{\@secondoftwo}%
\providecommand \href [0]{\begingroup \@sanitize@url \@href}%
\providecommand \@href[1]{\@@startlink{#1}\@@href}%
\providecommand \@@href[1]{\endgroup#1\@@endlink}%
\providecommand \@sanitize@url [0]{\catcode `\\12\catcode `\$12\catcode `\&12\catcode `\#12\catcode `\^12\catcode `\_12\catcode `\%12\relax}%
\providecommand \@@startlink[1]{}%
\providecommand \@@endlink[0]{}%
\providecommand \url  [0]{\begingroup\@sanitize@url \@url }%
\providecommand \@url [1]{\endgroup\@href {#1}{\urlprefix }}%
\providecommand \urlprefix  [0]{URL }%
\providecommand \Eprint [0]{\href }%
\providecommand \doibase [0]{https://doi.org/}%
\providecommand \selectlanguage [0]{\@gobble}%
\providecommand \bibinfo  [0]{\@secondoftwo}%
\providecommand \bibfield  [0]{\@secondoftwo}%
\providecommand \translation [1]{[#1]}%
\providecommand \BibitemOpen [0]{}%
\providecommand \bibitemStop [0]{}%
\providecommand \bibitemNoStop [0]{.\EOS\space}%
\providecommand \EOS [0]{\spacefactor3000\relax}%
\providecommand \BibitemShut  [1]{\csname bibitem#1\endcsname}%
\let\auto@bib@innerbib\@empty
\bibitem [{\citenamefont {Anderson}(1987)}]{anderson1987}%
  \BibitemOpen
  \bibfield  {author} {\bibinfo {author} {\bibfnamefont {P.~W.}\ \bibnamefont {Anderson}},\ }\bibfield  {title} {\bibinfo {title} {The resonating valence bond state in {L}a$_2${C}u{O}$_4$ and superconductivity},\ }\href {https://doi.org/10.1126/science.235.4793.1196} {\bibfield  {journal} {\bibinfo  {journal} {Science}\ }\textbf {\bibinfo {volume} {235}},\ \bibinfo {pages} {1196} (\bibinfo {year} {1987})}\BibitemShut {NoStop}%
\bibitem [{\citenamefont {Kivelson}\ \emph {et~al.}(1987)\citenamefont {Kivelson}, \citenamefont {Rokhsar},\ and\ \citenamefont {Sethna}}]{kivelson1987}%
  \BibitemOpen
  \bibfield  {author} {\bibinfo {author} {\bibfnamefont {S.~A.}\ \bibnamefont {Kivelson}}, \bibinfo {author} {\bibfnamefont {D.~S.}\ \bibnamefont {Rokhsar}},\ and\ \bibinfo {author} {\bibfnamefont {J.~P.}\ \bibnamefont {Sethna}},\ }\bibfield  {title} {\bibinfo {title} {Topology of the resonating valence-bond state: Solitons and high-${T}_c$ superconductivity},\ }\href {https://doi.org/10.1103/PhysRevB.35.8865} {\bibfield  {journal} {\bibinfo  {journal} {Phys. Rev. B}\ }\textbf {\bibinfo {volume} {35}},\ \bibinfo {pages} {8865} (\bibinfo {year} {1987})}\BibitemShut {NoStop}%
\bibitem [{\citenamefont {Rokhsar}\ and\ \citenamefont {Kivelson}(1988)}]{rokhsar1988}%
  \BibitemOpen
  \bibfield  {author} {\bibinfo {author} {\bibfnamefont {D.~S.}\ \bibnamefont {Rokhsar}}\ and\ \bibinfo {author} {\bibfnamefont {S.~A.}\ \bibnamefont {Kivelson}},\ }\bibfield  {title} {\bibinfo {title} {Superconductivity and the quantum hard-core dimer gas},\ }\href {https://doi.org/10.1103/PhysRevLett.61.2376} {\bibfield  {journal} {\bibinfo  {journal} {Phys. Rev. Lett.}\ }\textbf {\bibinfo {volume} {61}},\ \bibinfo {pages} {2376} (\bibinfo {year} {1988})}\BibitemShut {NoStop}%
\bibitem [{\citenamefont {Chayes}\ \emph {et~al.}(1989)\citenamefont {Chayes}, \citenamefont {Chayes},\ and\ \citenamefont {Kivelson}}]{Chayes1989}%
  \BibitemOpen
  \bibfield  {author} {\bibinfo {author} {\bibfnamefont {J.~T.}\ \bibnamefont {Chayes}}, \bibinfo {author} {\bibfnamefont {L.}~\bibnamefont {Chayes}},\ and\ \bibinfo {author} {\bibfnamefont {S.~A.}\ \bibnamefont {Kivelson}},\ }\bibfield  {title} {\bibinfo {title} {Valence bond ground states in a frustrated two-dimensional spin-1/2 {H}eisenberg antiferromagnet},\ }\href {https://doi.org/10.1007/BF01244017} {\bibfield  {journal} {\bibinfo  {journal} {Commun. Math. Phys.}\ }\textbf {\bibinfo {volume} {123}},\ \bibinfo {pages} {53} (\bibinfo {year} {1989})}\BibitemShut {NoStop}%
\bibitem [{\citenamefont {Sachdev}(1989)}]{sachdev1989}%
  \BibitemOpen
  \bibfield  {author} {\bibinfo {author} {\bibfnamefont {S.}~\bibnamefont {Sachdev}},\ }\bibfield  {title} {\bibinfo {title} {Spin-{P}eierls ground states of the quantum dimer model: A finite-size study},\ }\href {https://doi.org/10.1103/PhysRevB.40.5204} {\bibfield  {journal} {\bibinfo  {journal} {Phys. Rev. B}\ }\textbf {\bibinfo {volume} {40}},\ \bibinfo {pages} {5204} (\bibinfo {year} {1989})}\BibitemShut {NoStop}%
\bibitem [{\citenamefont {Chaubey}\ \emph {et~al.}(2026)\citenamefont {Chaubey}, \citenamefont {Moroz},\ and\ \citenamefont {Bhattacharjee}}]{chaubey2026quantumdimerspifluxtoric}%
  \BibitemOpen
  \bibfield  {author} {\bibinfo {author} {\bibfnamefont {A.}~\bibnamefont {Chaubey}}, \bibinfo {author} {\bibfnamefont {S.}~\bibnamefont {Moroz}},\ and\ \bibinfo {author} {\bibfnamefont {S.}~\bibnamefont {Bhattacharjee}},\ }\href {https://arxiv.org/abs/2603.23154} {\bibinfo {title} {From quantum dimers to the $\pi$-flux toric code via deconfined multicriticality}} (\bibinfo {year} {2026}),\ \Eprint {https://arxiv.org/abs/2603.23154} {arXiv:2603.23154 [cond-mat.str-el]} \BibitemShut {NoStop}%
\bibitem [{\citenamefont {Moessner}\ and\ \citenamefont {Sondhi}(2001)}]{moessner2001rvb}%
  \BibitemOpen
  \bibfield  {author} {\bibinfo {author} {\bibfnamefont {R.}~\bibnamefont {Moessner}}\ and\ \bibinfo {author} {\bibfnamefont {S.~L.}\ \bibnamefont {Sondhi}},\ }\bibfield  {title} {\bibinfo {title} {Resonating valence bond phase in the triangular lattice quantum dimer model},\ }\href {https://doi.org/10.1103/PhysRevLett.86.1881} {\bibfield  {journal} {\bibinfo  {journal} {Phys. Rev. Lett.}\ }\textbf {\bibinfo {volume} {86}},\ \bibinfo {pages} {1881} (\bibinfo {year} {2001})}\BibitemShut {NoStop}%
\bibitem [{\citenamefont {Moessner}\ and\ \citenamefont {Sondhi}(2003)}]{moessner2003}%
  \BibitemOpen
  \bibfield  {author} {\bibinfo {author} {\bibfnamefont {R.}~\bibnamefont {Moessner}}\ and\ \bibinfo {author} {\bibfnamefont {S.~L.}\ \bibnamefont {Sondhi}},\ }\bibfield  {title} {\bibinfo {title} {Three-dimensional resonating-valence-bond liquids and their excitations},\ }\href {https://doi.org/10.1103/PhysRevB.68.184512} {\bibfield  {journal} {\bibinfo  {journal} {Phys. Rev. B}\ }\textbf {\bibinfo {volume} {68}},\ \bibinfo {pages} {184512} (\bibinfo {year} {2003})}\BibitemShut {NoStop}%
\bibitem [{\citenamefont {Ardonne}\ \emph {et~al.}(2004)\citenamefont {Ardonne}, \citenamefont {Fendley},\ and\ \citenamefont {Fradkin}}]{ARDONNE2004493}%
  \BibitemOpen
  \bibfield  {author} {\bibinfo {author} {\bibfnamefont {E.}~\bibnamefont {Ardonne}}, \bibinfo {author} {\bibfnamefont {P.}~\bibnamefont {Fendley}},\ and\ \bibinfo {author} {\bibfnamefont {E.}~\bibnamefont {Fradkin}},\ }\bibfield  {title} {\bibinfo {title} {Topological order and conformal quantum critical points},\ }\href {https://doi.org/10.1016/j.aop.2004.01.004} {\bibfield  {journal} {\bibinfo  {journal} {Ann. Phys.}\ }\textbf {\bibinfo {volume} {310}},\ \bibinfo {pages} {493} (\bibinfo {year} {2004})}\BibitemShut {NoStop}%
\bibitem [{\citenamefont {Sachdev}(2008)}]{Sachdev2008}%
  \BibitemOpen
  \bibfield  {author} {\bibinfo {author} {\bibfnamefont {S.}~\bibnamefont {Sachdev}},\ }\bibfield  {title} {\bibinfo {title} {Quantum magnetism and criticality},\ }\href {https://doi.org/10.1038/nphys894} {\bibfield  {journal} {\bibinfo  {journal} {Nat. Phys.}\ }\textbf {\bibinfo {volume} {4}},\ \bibinfo {pages} {173} (\bibinfo {year} {2008})}\BibitemShut {NoStop}%
\bibitem [{\citenamefont {Sachdev}(2011)}]{Sachdev_2011}%
  \BibitemOpen
  \bibfield  {author} {\bibinfo {author} {\bibfnamefont {S.}~\bibnamefont {Sachdev}},\ }\href@noop {} {\emph {\bibinfo {title} {Quantum Phase Transitions}}},\ \bibinfo {edition} {2nd}\ ed.\ (\bibinfo  {publisher} {Cambridge University Press},\ \bibinfo {year} {2011})\BibitemShut {NoStop}%
\bibitem [{\citenamefont {Fradkin}(2013)}]{Fradkin_2013_dimer}%
  \BibitemOpen
  \bibfield  {author} {\bibinfo {author} {\bibfnamefont {E.}~\bibnamefont {Fradkin}},\ }\bibinfo {title} {Gauge theory, dimer models, and topological phases},\ in\ \href@noop {} {\emph {\bibinfo {booktitle} {Field Theories of Condensed Matter Physics}}}\ (\bibinfo  {publisher} {Cambridge University Press},\ \bibinfo {year} {2013})\ pp.\ \bibinfo {pages} {286--358}\BibitemShut {NoStop}%
\bibitem [{\citenamefont {Moessner}\ and\ \citenamefont {Raman}(2011)}]{moessner2011}%
  \BibitemOpen
  \bibfield  {author} {\bibinfo {author} {\bibfnamefont {R.}~\bibnamefont {Moessner}}\ and\ \bibinfo {author} {\bibfnamefont {K.~S.}\ \bibnamefont {Raman}},\ }\bibfield  {title} {\bibinfo {title} {Quantum dimer models},\ }in\ \href@noop {} {\emph {\bibinfo {booktitle} {Introduction to Frustrated Magnetism: Materials, Experiments, Theory}}}\ (\bibinfo  {publisher} {Springer},\ \bibinfo {year} {2011})\ pp.\ \bibinfo {pages} {437--479},\ \Eprint {https://arxiv.org/abs/0809.3051} {arXiv:0809.3051} \BibitemShut {NoStop}%
\bibitem [{\citenamefont {Samajdar}\ \emph {et~al.}(2021)\citenamefont {Samajdar}, \citenamefont {Ho}, \citenamefont {Pichler}, \citenamefont {Lukin},\ and\ \citenamefont {Sachdev}}]{samajdar2021rydberg}%
  \BibitemOpen
  \bibfield  {author} {\bibinfo {author} {\bibfnamefont {R.}~\bibnamefont {Samajdar}}, \bibinfo {author} {\bibfnamefont {W.~W.}\ \bibnamefont {Ho}}, \bibinfo {author} {\bibfnamefont {H.}~\bibnamefont {Pichler}}, \bibinfo {author} {\bibfnamefont {M.~D.}\ \bibnamefont {Lukin}},\ and\ \bibinfo {author} {\bibfnamefont {S.}~\bibnamefont {Sachdev}},\ }\bibfield  {title} {\bibinfo {title} {Quantum phases of {R}ydberg atoms on a {K}agome lattice},\ }\href {https://doi.org/10.1073/pnas.2015785118} {\bibfield  {journal} {\bibinfo  {journal} {Proc. Natl. Acad. Sci. U.S.A.}\ }\textbf {\bibinfo {volume} {118}},\ \bibinfo {pages} {e2015785118} (\bibinfo {year} {2021})} \BibitemShut {NoStop}%
\bibitem [{\citenamefont {Verresen}\ \emph {et~al.}(2021)\citenamefont {Verresen}, \citenamefont {Lukin},\ and\ \citenamefont {Vishwanath}}]{verresen2021prediction}%
  \BibitemOpen
  \bibfield  {author} {\bibinfo {author} {\bibfnamefont {R.}~\bibnamefont {Verresen}}, \bibinfo {author} {\bibfnamefont {M.~D.}\ \bibnamefont {Lukin}},\ and\ \bibinfo {author} {\bibfnamefont {A.}~\bibnamefont {Vishwanath}},\ }\bibfield  {title} {\bibinfo {title} {Prediction of toric code topological order from {R}ydberg blockade},\ }\href {https://doi.org/10.1103/PhysRevX.11.031005} {\bibfield  {journal} {\bibinfo  {journal} {Phys. Rev. X}\ }\textbf {\bibinfo {volume} {11}},\ \bibinfo {pages} {031005} (\bibinfo {year} {2021})}\BibitemShut {NoStop}%
\bibitem [{\citenamefont {Zeng}\ \emph {et~al.}(2025)\citenamefont {Zeng}, \citenamefont {Giudici},\ and\ \citenamefont {Pichler}}]{zeng2025dimer}%
  \BibitemOpen
  \bibfield  {author} {\bibinfo {author} {\bibfnamefont {Z.}~\bibnamefont {Zeng}}, \bibinfo {author} {\bibfnamefont {G.}~\bibnamefont {Giudici}},\ and\ \bibinfo {author} {\bibfnamefont {H.}~\bibnamefont {Pichler}},\ }\bibfield  {title} {\bibinfo {title} {Quantum dimer models with {R}ydberg gadgets},\ }\href {https://doi.org/10.1103/PhysRevResearch.7.L012006} {\bibfield  {journal} {\bibinfo  {journal} {Phys. Rev. Res.}\ }\textbf {\bibinfo {volume} {7}},\ \bibinfo {pages} {L012006} (\bibinfo {year} {2025})}\BibitemShut {NoStop}%
\bibitem [{\citenamefont {Rossi}\ and\ \citenamefont {Wolff}(1984)}]{ROSSI1984}%
  \BibitemOpen
  \bibfield  {author} {\bibinfo {author} {\bibfnamefont {P.}~\bibnamefont {Rossi}}\ and\ \bibinfo {author} {\bibfnamefont {U.}~\bibnamefont {Wolff}},\ }\bibfield  {title} {\bibinfo {title} {Lattice {QCD} with fermions at strong coupling: A dimer system},\ }\href {https://doi.org/10.1016/0550-3213(84)90589-3} {\bibfield  {journal} {\bibinfo  {journal} {Nucl. Phys. B}\ }\textbf {\bibinfo {volume} {248}},\ \bibinfo {pages} {105} (\bibinfo {year} {1984})}\BibitemShut {NoStop}%
\bibitem [{\citenamefont {Moessner}\ \emph {et~al.}(2001)\citenamefont {Moessner}, \citenamefont {Sondhi},\ and\ \citenamefont {Fradkin}}]{moessner2001ising}%
  \BibitemOpen
  \bibfield  {author} {\bibinfo {author} {\bibfnamefont {R.}~\bibnamefont {Moessner}}, \bibinfo {author} {\bibfnamefont {S.~L.}\ \bibnamefont {Sondhi}},\ and\ \bibinfo {author} {\bibfnamefont {E.}~\bibnamefont {Fradkin}},\ }\bibfield  {title} {\bibinfo {title} {Short-ranged resonating valence bond physics, quantum dimer models, and {I}sing gauge theories},\ }\href {https://doi.org/10.1103/PhysRevB.65.024504} {\bibfield  {journal} {\bibinfo  {journal} {Phys. Rev. B}\ }\textbf {\bibinfo {volume} {65}},\ \bibinfo {pages} {024504} (\bibinfo {year} {2001})}\BibitemShut {NoStop}%
\bibitem [{\citenamefont {Borla}\ \emph {et~al.}(2020)\citenamefont {Borla}, \citenamefont {Verresen}, \citenamefont {Grusdt},\ and\ \citenamefont {Moroz}}]{borla2020confined}%
  \BibitemOpen
  \bibfield  {author} {\bibinfo {author} {\bibfnamefont {U.}~\bibnamefont {Borla}}, \bibinfo {author} {\bibfnamefont {R.}~\bibnamefont {Verresen}}, \bibinfo {author} {\bibfnamefont {F.}~\bibnamefont {Grusdt}},\ and\ \bibinfo {author} {\bibfnamefont {S.}~\bibnamefont {Moroz}},\ }\bibfield  {title} {\bibinfo {title} {Confined phases of one-dimensional spinless fermions coupled to $\mathbb{{Z}}_{2}$ gauge theory},\ }\href {https://doi.org/10.1103/PhysRevLett.124.120503} {\bibfield  {journal} {\bibinfo  {journal} {Phys. Rev. Lett.}\ }\textbf {\bibinfo {volume} {124}},\ \bibinfo {pages} {120503} (\bibinfo {year} {2020})}\BibitemShut {NoStop}%
\bibitem [{\citenamefont {Borla}\ \emph {et~al.}(2022)\citenamefont {Borla}, \citenamefont {Jeevanesan}, \citenamefont {Pollmann},\ and\ \citenamefont {Moroz}}]{borla2022}%
  \BibitemOpen
  \bibfield  {author} {\bibinfo {author} {\bibfnamefont {U.}~\bibnamefont {Borla}}, \bibinfo {author} {\bibfnamefont {B.}~\bibnamefont {Jeevanesan}}, \bibinfo {author} {\bibfnamefont {F.}~\bibnamefont {Pollmann}},\ and\ \bibinfo {author} {\bibfnamefont {S.}~\bibnamefont {Moroz}},\ }\bibfield  {title} {\bibinfo {title} {Quantum phases of two-dimensional $\mathbb{Z}_{2}$ gauge theory coupled to single-component fermion matter},\ }\href {https://doi.org/10.1103/PhysRevB.105.075132} {\bibfield  {journal} {\bibinfo  {journal} {Phys. Rev. B}\ }\textbf {\bibinfo {volume} {105}},\ \bibinfo {pages} {075132} (\bibinfo {year} {2022})}\BibitemShut {NoStop}%
\bibitem [{\citenamefont {Borla}\ \emph {et~al.}(2026)\citenamefont {Borla}, \citenamefont {De},\ and\ \citenamefont {Gazit}}]{borla2026odd}%
  \BibitemOpen
  \bibfield  {author} {\bibinfo {author} {\bibfnamefont {U.}~\bibnamefont {Borla}}, \bibinfo {author} {\bibfnamefont {A.}~\bibnamefont {De}},\ and\ \bibinfo {author} {\bibfnamefont {S.}~\bibnamefont {Gazit}},\ }\bibfield  {title} {\bibinfo {title} {Odd toric code in a tilted field: {H}iggs-confinement multicriticality, spontaneous self-duality symmetry breaking, and valence bond solids},\ }\href {https://doi.org/10.1103/wbvd-z9s4} {\bibfield  {journal} {\bibinfo  {journal} {Phys. Rev. B}\ }\textbf {\bibinfo {volume} {113}},\ \bibinfo {pages} {245134} (\bibinfo {year} {2026})}\BibitemShut {NoStop}%
\bibitem [{\citenamefont {Sachdev}(2023)}]{Sachdev_2023}%
  \BibitemOpen
  \bibfield  {author} {\bibinfo {author} {\bibfnamefont {S.}~\bibnamefont {Sachdev}},\ }\href@noop {} {\emph {\bibinfo {title} {Quantum Phases of Matter}}}\ (\bibinfo  {publisher} {Cambridge University Press},\ \bibinfo {year} {2023})\BibitemShut {NoStop}%
\bibitem [{\citenamefont {Pollmann}\ \emph {et~al.}(2011)\citenamefont {Pollmann}, \citenamefont {Betouras}, \citenamefont {Shtengel},\ and\ \citenamefont {Fulde}}]{pollmann2011fermionic}%
  \BibitemOpen
  \bibfield  {author} {\bibinfo {author} {\bibfnamefont {F.}~\bibnamefont {Pollmann}}, \bibinfo {author} {\bibfnamefont {J.~J.}\ \bibnamefont {Betouras}}, \bibinfo {author} {\bibfnamefont {K.}~\bibnamefont {Shtengel}},\ and\ \bibinfo {author} {\bibfnamefont {P.}~\bibnamefont {Fulde}},\ }\bibfield  {title} {\bibinfo {title} {Fermionic quantum dimer and fully-packed loop models on the square lattice},\ }\href {https://doi.org/10.1103/PhysRevB.83.155117} {\bibfield  {journal} {\bibinfo  {journal} {Phys. Rev. B}\ }\textbf {\bibinfo {volume} {83}},\ \bibinfo {pages} {155117} (\bibinfo {year} {2011})}\BibitemShut {NoStop}%
\bibitem [{\citenamefont {Punk}\ \emph {et~al.}(2015)\citenamefont {Punk}, \citenamefont {Allais},\ and\ \citenamefont {Sachdev}}]{punk2015}%
  \BibitemOpen
  \bibfield  {author} {\bibinfo {author} {\bibfnamefont {M.}~\bibnamefont {Punk}}, \bibinfo {author} {\bibfnamefont {A.}~\bibnamefont {Allais}},\ and\ \bibinfo {author} {\bibfnamefont {S.}~\bibnamefont {Sachdev}},\ }\bibfield  {title} {\bibinfo {title} {Quantum dimer model for the pseudogap metal},\ }\href {https://doi.org/10.1073/pnas.1512206112} {\bibfield  {journal} {\bibinfo  {journal} {Proc. Natl. Acad. Sci. U.S.A.}\ }\textbf {\bibinfo {volume} {112}},\ \bibinfo {pages} {9552} (\bibinfo {year} {2015})} \BibitemShut {NoStop}%
\bibitem [{\citenamefont {Feldmeier}\ \emph {et~al.}(2018)\citenamefont {Feldmeier}, \citenamefont {Huber},\ and\ \citenamefont {Punk}}]{feldmeier2018exact}%
  \BibitemOpen
  \bibfield  {author} {\bibinfo {author} {\bibfnamefont {J.}~\bibnamefont {Feldmeier}}, \bibinfo {author} {\bibfnamefont {S.}~\bibnamefont {Huber}},\ and\ \bibinfo {author} {\bibfnamefont {M.}~\bibnamefont {Punk}},\ }\bibfield  {title} {\bibinfo {title} {Exact solution of a two-species quantum dimer model for pseudogap metals},\ }\href {https://doi.org/10.1103/PhysRevLett.120.187001} {\bibfield  {journal} {\bibinfo  {journal} {Phys. Rev. Lett.}\ }\textbf {\bibinfo {volume} {120}},\ \bibinfo {pages} {187001} (\bibinfo {year} {2018})}\BibitemShut {NoStop}%
\bibitem [{\citenamefont {Wegner}(1971)}]{wegner1971}%
  \BibitemOpen
  \bibfield  {author} {\bibinfo {author} {\bibfnamefont {F.~J.}\ \bibnamefont {Wegner}},\ }%
  \bibfield  {title} {\bibinfo {title} {Duality in generalized {I}sing models and phase transitions without local order parameters},\ }%
  \href {https://doi.org/10.1063/1.1665530}%
  {\bibfield  {journal} {\bibinfo {journal} {J. Math. Phys.}\ }%
  \textbf {\bibinfo {volume} {12}},\ \bibinfo {pages} {2259} (\bibinfo {year} {1971})}%
  \BibitemShut {NoStop}%
\bibitem [{\citenamefont {Fradkin}\ and\ \citenamefont {Shenker}(1979)}]{fs1979}%
  \BibitemOpen
  \bibfield  {author} {\bibinfo {author} {\bibfnamefont {E.}~\bibnamefont {Fradkin}}\ and\ \bibinfo {author} {\bibfnamefont {S.~H.}\ \bibnamefont {Shenker}},\ }\bibfield  {title} {\bibinfo {title} {Phase diagrams of lattice gauge theories with {H}iggs fields},\ }\href {https://doi.org/10.1103/PhysRevD.19.3682} {\bibfield  {journal} {\bibinfo  {journal} {Phys. Rev. D}\ }\textbf {\bibinfo {volume} {19}},\ \bibinfo {pages} {3682} (\bibinfo {year} {1979})}\BibitemShut {NoStop}%
\bibitem [{\citenamefont {Senthil}\ and\ \citenamefont {Fisher}(2000)}]{senthil2000}%
  \BibitemOpen
  \bibfield  {author} {\bibinfo {author} {\bibfnamefont {T.}~\bibnamefont {Senthil}}\ and\ \bibinfo {author} {\bibfnamefont {M.~P.~A.}\ \bibnamefont {Fisher}},\ }\bibfield  {title} {\bibinfo {title} {$\mathbb{{Z}}_2$ gauge theory of electron fractionalization in strongly correlated systems},\ }\href {https://doi.org/10.1103/PhysRevB.62.7850} {\bibfield  {journal} {\bibinfo  {journal} {Phys. Rev. B}\ }\textbf {\bibinfo {volume} {62}},\ \bibinfo {pages} {7850} (\bibinfo {year} {2000})}\BibitemShut {NoStop}%
\bibitem [{\citenamefont {Assaad}\ and\ \citenamefont {Grover}(2016)}]{assaad2016}%
  \BibitemOpen
  \bibfield  {author} {\bibinfo {author} {\bibfnamefont {F.~F.}\ \bibnamefont {Assaad}}\ and\ \bibinfo {author} {\bibfnamefont {T.}~\bibnamefont {Grover}},\ }\bibfield  {title} {\bibinfo {title} {Simple fermionic model of deconfined phases and phase transitions},\ }\href {https://doi.org/10.1103/PhysRevX.6.041049} {\bibfield  {journal} {\bibinfo  {journal} {Phys. Rev. X}\ }\textbf {\bibinfo {volume} {6}},\ \bibinfo {pages} {041049} (\bibinfo {year} {2016})}\BibitemShut {NoStop}%
\bibitem [{\citenamefont {Gazit}\ \emph {et~al.}(2017)\citenamefont {Gazit}, \citenamefont {Randeria},\ and\ \citenamefont {Vishwanath}}]{gazit2017}%
  \BibitemOpen
  \bibfield  {author} {\bibinfo {author} {\bibfnamefont {S.}~\bibnamefont {Gazit}}, \bibinfo {author} {\bibfnamefont {M.}~\bibnamefont {Randeria}},\ and\ \bibinfo {author} {\bibfnamefont {A.}~\bibnamefont {Vishwanath}},\ }\bibfield  {title} {\bibinfo {title} {Emergent {D}irac fermions and broken symmetries in confined and deconfined phases of $\mathbb{{Z}}_2$ gauge theories},\ }\href {https://doi.org/10.1038/nphys4028} {\bibfield  {journal} {\bibinfo  {journal} {Nat. Phys.}\ }\textbf {\bibinfo {volume} {13}},\ \bibinfo {pages} {484} (\bibinfo {year} {2017})}\BibitemShut {NoStop}%
\bibitem [{\citenamefont {Kebri{\v{c}}}\ \emph {et~al.}(2021)\citenamefont {Kebri{\v{c}}}, \citenamefont {Barbiero}, \citenamefont {Reinmoser}, \citenamefont {Schollw\"ock},\ and\ \citenamefont {Grusdt}}]{kebric2021}%
  \BibitemOpen
  \bibfield  {author} {\bibinfo {author} {\bibfnamefont {M.}~\bibnamefont {Kebri{\v{c}}}}, \bibinfo {author} {\bibfnamefont {L.}~\bibnamefont {Barbiero}}, \bibinfo {author} {\bibfnamefont {C.}~\bibnamefont {Reinmoser}}, \bibinfo {author} {\bibfnamefont {U.}~\bibnamefont {Schollw\"ock}},\ and\ \bibinfo {author} {\bibfnamefont {F.}~\bibnamefont {Grusdt}},\ }\bibfield  {title} {\bibinfo {title} {Confinement and {M}ott transitions of dynamical charges in one-dimensional lattice gauge theories},\ }\href {https://doi.org/10.1103/PhysRevLett.127.167203} {\bibfield  {journal} {\bibinfo  {journal} {Phys. Rev. Lett.}\ }\textbf {\bibinfo {volume} {127}},\ \bibinfo {pages} {167203} (\bibinfo {year} {2021})}\BibitemShut {NoStop}%
\bibitem [{\citenamefont {Borla}\ \emph {et~al.}(2024)\citenamefont {Borla}, \citenamefont {Gazit},\ and\ \citenamefont {Moroz}}]{borla2024}%
  \BibitemOpen
  \bibfield  {author} {\bibinfo {author} {\bibfnamefont {U.}~\bibnamefont {Borla}}, \bibinfo {author} {\bibfnamefont {S.}~\bibnamefont {Gazit}},\ and\ \bibinfo {author} {\bibfnamefont {S.}~\bibnamefont {Moroz}},\ }\bibfield  {title} {\bibinfo {title} {Deconfined quantum criticality in {I}sing gauge theory entangled with single-component fermions},\ }\href {https://doi.org/10.1103/PhysRevB.110.L201110} {\bibfield  {journal} {\bibinfo  {journal} {Phys. Rev. B}\ }\textbf {\bibinfo {volume} {110}},\ \bibinfo {pages} {L201110} (\bibinfo {year} {2024})}\BibitemShut {NoStop}%
\bibitem [{\citenamefont {Kebri{\v{c}}}\ \emph {et~al.}(2026)\citenamefont {Kebri{\v{c}}}, \citenamefont {D{\"o}schl}, \citenamefont {Borla}, \citenamefont {Halimeh}, \citenamefont {Schollw{\"o}ck}, \citenamefont {Bohrdt},\ and\ \citenamefont {Grusdt}}]{kebric2026}%
  \BibitemOpen
  \bibfield  {author} {\bibinfo {author} {\bibfnamefont {M.}~\bibnamefont {Kebri{\v{c}}}}, \bibinfo {author} {\bibfnamefont {F.}~\bibnamefont {D{\"o}schl}}, \bibinfo {author} {\bibfnamefont {U.}~\bibnamefont {Borla}}, \bibinfo {author} {\bibfnamefont {J.~C.}\ \bibnamefont {Halimeh}}, \bibinfo {author} {\bibfnamefont {U.}~\bibnamefont {Schollw{\"o}ck}}, \bibinfo {author} {\bibfnamefont {A.}~\bibnamefont {Bohrdt}},\ and\ \bibinfo {author} {\bibfnamefont {F.}~\bibnamefont {Grusdt}},\ }\href {https://arxiv.org/abs/2602.13192} {\bibinfo {title} {Matter-induced plaquette terms in a $\mathbb{Z}_2$ lattice gauge theory}} (\bibinfo {year} {2026}),\ \Eprint {https://arxiv.org/abs/2602.13192} {arXiv:2602.13192 [cond-mat.quant-gas]} \BibitemShut {NoStop}%
\bibitem [{\citenamefont {Homeier}\ \emph {et~al.}(2023)\citenamefont {Homeier}, \citenamefont {Bohrdt}, \citenamefont {Linsel}, \citenamefont {Demler}, \citenamefont {Halimeh},\ and\ \citenamefont {Grusdt}}]{Homeier2023}%
  \BibitemOpen
  \bibfield  {author} {\bibinfo {author} {\bibfnamefont {L.}~\bibnamefont {Homeier}}, \bibinfo {author} {\bibfnamefont {A.}~\bibnamefont {Bohrdt}}, \bibinfo {author} {\bibfnamefont {S.}~\bibnamefont {Linsel}}, \bibinfo {author} {\bibfnamefont {E.}~\bibnamefont {Demler}}, \bibinfo {author} {\bibfnamefont {J.~C.}\ \bibnamefont {Halimeh}},\ and\ \bibinfo {author} {\bibfnamefont {F.}~\bibnamefont {Grusdt}},\ }\bibfield  {title} {\bibinfo {title} {Realistic scheme for quantum simulation of $\mathbb{{Z}}_2$ lattice gauge theories with dynamical matter in (2{\thinspace}+{\thinspace}1)d},\ }\href {https://doi.org/10.1038/s42005-023-01237-6} {\bibfield  {journal} {\bibinfo  {journal} {Commun. Phys.}\ }\textbf {\bibinfo {volume} {6}},\ \bibinfo {pages} {127} (\bibinfo {year} {2023})}\BibitemShut {NoStop}%
\bibitem [{\citenamefont {Senthil}\ \emph {et~al.}(2004{\natexlab{a}})\citenamefont {Senthil}, \citenamefont {Vishwanath}, \citenamefont {Balents}, \citenamefont {Sachdev},\ and\ \citenamefont {Fisher}}]{senthil2004}%
  \BibitemOpen
  \bibfield  {author} {\bibinfo {author} {\bibfnamefont {T.}~\bibnamefont {Senthil}}, \bibinfo {author} {\bibfnamefont {A.}~\bibnamefont {Vishwanath}}, \bibinfo {author} {\bibfnamefont {L.}~\bibnamefont {Balents}}, \bibinfo {author} {\bibfnamefont {S.}~\bibnamefont {Sachdev}},\ and\ \bibinfo {author} {\bibfnamefont {M.~P.~A.}\ \bibnamefont {Fisher}},\ }\bibfield  {title} {\bibinfo {title} {Deconfined quantum critical points},\ }\href {https://doi.org/10.1126/science.1091806} {\bibfield  {journal} {\bibinfo  {journal} {Science}\ }\textbf {\bibinfo {volume} {303}},\ \bibinfo {pages} {1490} (\bibinfo {year} {2004}{\natexlab{a}})} \BibitemShut {NoStop}%
\bibitem [{\citenamefont {Senthil}\ \emph {et~al.}(2004{\natexlab{b}})\citenamefont {Senthil}, \citenamefont {Balents}, \citenamefont {Sachdev}, \citenamefont {Vishwanath},\ and\ \citenamefont {Fisher}}]{senthil2004review}%
  \BibitemOpen
  \bibfield  {author} {\bibinfo {author} {\bibfnamefont {T.}~\bibnamefont {Senthil}}, \bibinfo {author} {\bibfnamefont {L.}~\bibnamefont {Balents}}, \bibinfo {author} {\bibfnamefont {S.}~\bibnamefont {Sachdev}}, \bibinfo {author} {\bibfnamefont {A.}~\bibnamefont {Vishwanath}},\ and\ \bibinfo {author} {\bibfnamefont {M.~P.~A.}\ \bibnamefont {Fisher}},\ }\bibfield  {title} {\bibinfo {title} {Quantum criticality beyond the {L}andau--{G}inzburg--{W}ilson paradigm},\ }\href {https://doi.org/10.1103/PhysRevB.70.144407} {\bibfield  {journal} {\bibinfo  {journal} {Phys. Rev. B}\ }\textbf {\bibinfo {volume} {70}},\ \bibinfo {pages} {144407} (\bibinfo {year} {2004}{\natexlab{b}})}\BibitemShut {NoStop}%
\bibitem [{\citenamefont {Levin}\ and\ \citenamefont {Senthil}(2004)}]{levin2004}%
  \BibitemOpen
  \bibfield  {author} {\bibinfo {author} {\bibfnamefont {M.}~\bibnamefont {Levin}}\ and\ \bibinfo {author} {\bibfnamefont {T.}~\bibnamefont {Senthil}},\ }\bibfield  {title} {\bibinfo {title} {Deconfined quantum criticality and {N}\'eel order via dimer disorder},\ }\href {https://doi.org/10.1103/PhysRevB.70.220403} {\bibfield  {journal} {\bibinfo  {journal} {Phys. Rev. B}\ }\textbf {\bibinfo {volume} {70}},\ \bibinfo {pages} {220403(R)} (\bibinfo {year} {2004})}\BibitemShut {NoStop}%
\bibitem [{\citenamefont {Wang}\ \emph {et~al.}(2017)\citenamefont {Wang}, \citenamefont {Nahum}, \citenamefont {Metlitski}, \citenamefont {Xu},\ and\ \citenamefont {Senthil}}]{wang2017deconfined}%
  \BibitemOpen
  \bibfield  {author} {\bibinfo {author} {\bibfnamefont {C.}~\bibnamefont {Wang}}, \bibinfo {author} {\bibfnamefont {A.}~\bibnamefont {Nahum}}, \bibinfo {author} {\bibfnamefont {M.~A.}\ \bibnamefont {Metlitski}}, \bibinfo {author} {\bibfnamefont {C.}~\bibnamefont {Xu}},\ and\ \bibinfo {author} {\bibfnamefont {T.}~\bibnamefont {Senthil}},\ }\bibfield  {title} {\bibinfo {title} {Deconfined quantum critical points: Symmetries and dualities},\ }\href {https://doi.org/10.1103/PhysRevX.7.031051} {\bibfield  {journal} {\bibinfo  {journal} {Phys. Rev. X}\ }\textbf {\bibinfo {volume} {7}},\ \bibinfo {pages} {031051} (\bibinfo {year} {2017})}\BibitemShut {NoStop}%
\bibitem [{\citenamefont {Senthil}(2023)}]{senthil2023deconfinedquantumcriticalpoints}%
  \BibitemOpen
  \bibfield  {author} {\bibinfo {author} {\bibfnamefont {T.}~\bibnamefont {Senthil}},\ }\href {https://arxiv.org/abs/2306.12638} {\bibinfo {title} {Deconfined quantum critical points: a review}} (\bibinfo {year} {2023}),\ \Eprint {https://arxiv.org/abs/2306.12638} {arXiv:2306.12638 [cond-mat.str-el]} \BibitemShut {NoStop}%
\bibitem [{\citenamefont {Aidelsburger}\ \emph {et~al.}(2021)\citenamefont {Aidelsburger}, \citenamefont {Barbiero}, \citenamefont {Bermudez}, \citenamefont {Chanda}, \citenamefont {Dauphin}, \citenamefont {Gonz{\'a}lez-Cuadra}, \citenamefont {Grzybowski}, \citenamefont {Hands}, \citenamefont {Jendrzejewski}, \citenamefont {J{\"u}nemann}, \citenamefont {Juzeli{\={u}}nas}, \citenamefont {Kasper}, \citenamefont {Piga}, \citenamefont {Ran}, \citenamefont {Rizzi}, \citenamefont {Sierra}, \citenamefont {Tagliacozzo}, \citenamefont {Tirrito}, \citenamefont {Zache}, \citenamefont {Zakrzewski}, \citenamefont {Zohar},\ and\ \citenamefont {Lewenstein}}]{aidelsburger2021}%
  \BibitemOpen
  \bibfield  {author} {\bibinfo {author} {\bibfnamefont {M.}~\bibnamefont {Aidelsburger}}, \bibinfo {author} {\bibfnamefont {L.}~\bibnamefont {Barbiero}}, \bibinfo {author} {\bibfnamefont {A.}~\bibnamefont {Bermudez}}, \bibinfo {author} {\bibfnamefont {T.}~\bibnamefont {Chanda}}, \bibinfo {author} {\bibfnamefont {A.}~\bibnamefont {Dauphin}}, \bibinfo {author} {\bibfnamefont {D.}~\bibnamefont {Gonz{\'a}lez-Cuadra}}, \bibinfo {author} {\bibfnamefont {P.~R.}\ \bibnamefont {Grzybowski}}, \bibinfo {author} {\bibfnamefont {S.}~\bibnamefont {Hands}}, \bibinfo {author} {\bibfnamefont {F.}~\bibnamefont {Jendrzejewski}}, \bibinfo {author} {\bibfnamefont {J.}~\bibnamefont {J{\"u}nemann}}, \bibinfo {author} {\bibfnamefont {G.}~\bibnamefont {Juzeli{\={u}}nas}}, \bibinfo {author} {\bibfnamefont {V.}~\bibnamefont {Kasper}}, \bibinfo {author} {\bibfnamefont {A.}~\bibnamefont {Piga}}, \bibinfo {author} {\bibfnamefont {S.-J.}\ \bibnamefont {Ran}}, \bibinfo {author} {\bibfnamefont {M.}~\bibnamefont {Rizzi}}, \bibinfo {author}
  {\bibfnamefont {G.}~\bibnamefont {Sierra}}, \bibinfo {author} {\bibfnamefont {L.}~\bibnamefont {Tagliacozzo}}, \bibinfo {author} {\bibfnamefont {E.}~\bibnamefont {Tirrito}}, \bibinfo {author} {\bibfnamefont {T.~V.}\ \bibnamefont {Zache}}, \bibinfo {author} {\bibfnamefont {J.}~\bibnamefont {Zakrzewski}}, \bibinfo {author} {\bibfnamefont {E.}~\bibnamefont {Zohar}},\ and\ \bibinfo {author} {\bibfnamefont {M.}~\bibnamefont {Lewenstein}},\ }\bibfield  {title} {\bibinfo {title} {Cold atoms meet lattice gauge theory},\ }\href {https://doi.org/10.1098/rsta.2021.0064} {\bibfield  {journal} {\bibinfo  {journal} {Phil. Trans. R. Soc. A}\ }\textbf {\bibinfo {volume} {380}},\ \bibinfo {pages} {20210064} (\bibinfo {year} {2021})} \BibitemShut {NoStop}%
\bibitem [{\citenamefont {Zohar}(2021)}]{zohar2021}%
  \BibitemOpen
  \bibfield  {author} {\bibinfo {author} {\bibfnamefont {E.}~\bibnamefont {Zohar}},\ }\bibfield  {title} {\bibinfo {title} {Quantum simulation of lattice gauge theories in more than one space dimension---requirements, challenges and methods},\ }\href {https://doi.org/10.1098/rsta.2021.0069} {\bibfield  {journal} {\bibinfo  {journal} {Phil. Trans. R. Soc. A}\ }\textbf {\bibinfo {volume} {380}},\ \bibinfo {pages} {20210069} (\bibinfo {year} {2021})} \BibitemShut {NoStop}%
\bibitem [{\citenamefont {Klco}\ \emph {et~al.}(2022)\citenamefont {Klco}, \citenamefont {Roggero},\ and\ \citenamefont {Savage}}]{Klco_2022}%
  \BibitemOpen
  \bibfield  {author} {\bibinfo {author} {\bibfnamefont {N.}~\bibnamefont {Klco}}, \bibinfo {author} {\bibfnamefont {A.}~\bibnamefont {Roggero}},\ and\ \bibinfo {author} {\bibfnamefont {M.~J.}\ \bibnamefont {Savage}},\ }\bibfield  {title} {\bibinfo {title} {Standard model physics and the digital quantum revolution: thoughts about the interface},\ }\href {https://doi.org/10.1088/1361-6633/ac58a4} {\bibfield  {journal} {\bibinfo  {journal} {Rep. Prog. Phys.}\ }\textbf {\bibinfo {volume} {85}},\ \bibinfo {pages} {064301} (\bibinfo {year} {2022})}\BibitemShut {NoStop}%
\bibitem [{\citenamefont {Irmejs}\ \emph {et~al.}(2023)\citenamefont {Irmejs}, \citenamefont {Ba\~nuls},\ and\ \citenamefont {Cirac}}]{Irmejs2023minimal}%
  \BibitemOpen
  \bibfield  {author} {\bibinfo {author} {\bibfnamefont {R.}~\bibnamefont {Irmejs}}, \bibinfo {author} {\bibfnamefont {M.-C.}\ \bibnamefont {Ba\~nuls}},\ and\ \bibinfo {author} {\bibfnamefont {J.~I.}\ \bibnamefont {Cirac}},\ }\bibfield  {title} {\bibinfo {title} {Quantum simulation of $\mathbb{{Z}}_{2}$ lattice gauge theory with minimal resources},\ }\href {https://doi.org/10.1103/PhysRevD.108.074503} {\bibfield  {journal} {\bibinfo  {journal} {Phys. Rev. D}\ }\textbf {\bibinfo {volume} {108}},\ \bibinfo {pages} {074503} (\bibinfo {year} {2023})}\BibitemShut {NoStop}%
\bibitem [{\citenamefont {Bauer}\ \emph {et~al.}(2023{\natexlab{a}})\citenamefont {Bauer}, \citenamefont {Davoudi}, \citenamefont {Klco},\ and\ \citenamefont {Savage}}]{Bauer2023}%
  \BibitemOpen
  \bibfield  {author} {\bibinfo {author} {\bibfnamefont {C.~W.}\ \bibnamefont {Bauer}}, \bibinfo {author} {\bibfnamefont {Z.}~\bibnamefont {Davoudi}}, \bibinfo {author} {\bibfnamefont {N.}~\bibnamefont {Klco}},\ and\ \bibinfo {author} {\bibfnamefont {M.~J.}\ \bibnamefont {Savage}},\ }\bibfield  {title} {\bibinfo {title} {Quantum simulation of fundamental particles and forces},\ }\href {https://doi.org/10.1038/s42254-023-00599-8} {\bibfield  {journal} {\bibinfo  {journal} {Nat. Rev. Phys.}\ }\textbf {\bibinfo {volume} {5}},\ \bibinfo {pages} {420} (\bibinfo {year} {2023}{\natexlab{a}})}\BibitemShut {NoStop}%
\bibitem [{\citenamefont {Bauer}\ \emph {et~al.}(2023{\natexlab{b}})\citenamefont {Bauer}, \citenamefont {Davoudi}, \citenamefont {Balantekin}, \citenamefont {Bhattacharya}, \citenamefont {Carena}, \citenamefont {de~Jong}, \citenamefont {Draper}, \citenamefont {El-Khadra}, \citenamefont {Gemelke}, \citenamefont {Hanada}, \citenamefont {Kharzeev}, \citenamefont {Lamm}, \citenamefont {Li}, \citenamefont {Liu}, \citenamefont {Lukin}, \citenamefont {Meurice}, \citenamefont {Monroe}, \citenamefont {Nachman}, \citenamefont {Pagano}, \citenamefont {Preskill}, \citenamefont {Rinaldi}, \citenamefont {Roggero}, \citenamefont {Santiago}, \citenamefont {Savage}, \citenamefont {Siddiqi}, \citenamefont {Siopsis}, \citenamefont {Van~Zanten}, \citenamefont {Wiebe}, \citenamefont {Yamauchi}, \citenamefont {Yeter-Aydeniz},\ and\ \citenamefont {Zorzetti}}]{bauer2023hep}%
  \BibitemOpen
  \bibfield  {author} {\bibinfo {author} {\bibfnamefont {C.~W.}\ \bibnamefont {Bauer}}, \bibinfo {author} {\bibfnamefont {Z.}~\bibnamefont {Davoudi}}, \bibinfo {author} {\bibfnamefont {A.~B.}\ \bibnamefont {Balantekin}}, \bibinfo {author} {\bibfnamefont {T.}~\bibnamefont {Bhattacharya}}, \bibinfo {author} {\bibfnamefont {M.}~\bibnamefont {Carena}}, \bibinfo {author} {\bibfnamefont {W.~A.}\ \bibnamefont {de~Jong}}, \bibinfo {author} {\bibfnamefont {P.}~\bibnamefont {Draper}}, \bibinfo {author} {\bibfnamefont {A.}~\bibnamefont {El-Khadra}}, \bibinfo {author} {\bibfnamefont {N.}~\bibnamefont {Gemelke}}, \bibinfo {author} {\bibfnamefont {M.}~\bibnamefont {Hanada}}, \bibinfo {author} {\bibfnamefont {D.}~\bibnamefont {Kharzeev}}, \bibinfo {author} {\bibfnamefont {H.}~\bibnamefont {Lamm}}, \bibinfo {author} {\bibfnamefont {Y.-Y.}\ \bibnamefont {Li}}, \bibinfo {author} {\bibfnamefont {J.}~\bibnamefont {Liu}}, \bibinfo {author} {\bibfnamefont {M.}~\bibnamefont {Lukin}}, \bibinfo {author} {\bibfnamefont
  {Y.}~\bibnamefont {Meurice}}, \bibinfo {author} {\bibfnamefont {C.}~\bibnamefont {Monroe}}, \bibinfo {author} {\bibfnamefont {B.}~\bibnamefont {Nachman}}, \bibinfo {author} {\bibfnamefont {G.}~\bibnamefont {Pagano}}, \bibinfo {author} {\bibfnamefont {J.}~\bibnamefont {Preskill}}, \bibinfo {author} {\bibfnamefont {E.}~\bibnamefont {Rinaldi}}, \bibinfo {author} {\bibfnamefont {A.}~\bibnamefont {Roggero}}, \bibinfo {author} {\bibfnamefont {D.~I.}\ \bibnamefont {Santiago}}, \bibinfo {author} {\bibfnamefont {M.~J.}\ \bibnamefont {Savage}}, \bibinfo {author} {\bibfnamefont {I.}~\bibnamefont {Siddiqi}}, \bibinfo {author} {\bibfnamefont {G.}~\bibnamefont {Siopsis}}, \bibinfo {author} {\bibfnamefont {D.}~\bibnamefont {Van~Zanten}}, \bibinfo {author} {\bibfnamefont {N.}~\bibnamefont {Wiebe}}, \bibinfo {author} {\bibfnamefont {Y.}~\bibnamefont {Yamauchi}}, \bibinfo {author} {\bibfnamefont {K.}~\bibnamefont {Yeter-Aydeniz}},\ and\ \bibinfo {author} {\bibfnamefont {S.}~\bibnamefont {Zorzetti}},\ }\bibfield  {title}
  {\bibinfo {title} {Quantum simulation for high-energy physics},\ }\href {https://doi.org/10.1103/PRXQuantum.4.027001} {\bibfield  {journal} {\bibinfo  {journal} {PRX Quantum}\ }\textbf {\bibinfo {volume} {4}},\ \bibinfo {pages} {027001} (\bibinfo {year} {2023}{\natexlab{b}})}\BibitemShut {NoStop}%
\bibitem [{\citenamefont {Di~Meglio}\ \emph {et~al.}(2024)\citenamefont {Di~Meglio} \emph {et~al.}}]{dimeglio2024}%
  \BibitemOpen
  \bibfield  {author} {\bibinfo {author} {\bibfnamefont {A.}~\bibnamefont {Di~Meglio}} \emph {et~al.},\ }\bibfield  {title} {\bibinfo {title} {Quantum computing for high-energy physics: State of the art and challenges},\ }\href {https://doi.org/10.1103/PRXQuantum.5.037001} {\bibfield  {journal} {\bibinfo  {journal} {PRX Quantum}\ }\textbf {\bibinfo {volume} {5}},\ \bibinfo {pages} {037001} (\bibinfo {year} {2024})}\BibitemShut {NoStop}%
\bibitem [{\citenamefont {Mildenberger}\ \emph {et~al.}(2025)\citenamefont {Mildenberger}, \citenamefont {Mruczkiewicz}, \citenamefont {Halimeh}, \citenamefont {Jiang},\ and\ \citenamefont {Hauke}}]{mildenberger2025}%
  \BibitemOpen
  \bibfield  {author} {\bibinfo {author} {\bibfnamefont {J.}~\bibnamefont {Mildenberger}}, \bibinfo {author} {\bibfnamefont {W.}~\bibnamefont {Mruczkiewicz}}, \bibinfo {author} {\bibfnamefont {J.~C.}\ \bibnamefont {Halimeh}}, \bibinfo {author} {\bibfnamefont {Z.}~\bibnamefont {Jiang}},\ and\ \bibinfo {author} {\bibfnamefont {P.}~\bibnamefont {Hauke}},\ }\bibfield  {title} {\bibinfo {title} {Confinement in a $\mathbb{Z}_2$ lattice gauge theory on a quantum computer},\ }\href {https://doi.org/10.1038/s41567-024-02723-6} {\bibfield  {journal} {\bibinfo  {journal} {Nat. Phys.}\ }\textbf {\bibinfo {volume} {21}},\ \bibinfo {pages} {312} (\bibinfo {year} {2025})}\BibitemShut {NoStop}%
\bibitem [{\citenamefont {Halimeh}\ \emph {et~al.}(2025{\natexlab{a}})\citenamefont {Halimeh}, \citenamefont {Aidelsburger}, \citenamefont {Grusdt}, \citenamefont {Hauke},\ and\ \citenamefont {Yang}}]{Halimeh2025}%
  \BibitemOpen
  \bibfield  {author} {\bibinfo {author} {\bibfnamefont {J.~C.}\ \bibnamefont {Halimeh}}, \bibinfo {author} {\bibfnamefont {M.}~\bibnamefont {Aidelsburger}}, \bibinfo {author} {\bibfnamefont {F.}~\bibnamefont {Grusdt}}, \bibinfo {author} {\bibfnamefont {P.}~\bibnamefont {Hauke}},\ and\ \bibinfo {author} {\bibfnamefont {B.}~\bibnamefont {Yang}},\ }\bibfield  {title} {\bibinfo {title} {Cold-atom quantum simulators of gauge theories},\ }\href {https://doi.org/10.1038/s41567-024-02721-8} {\bibfield  {journal} {\bibinfo  {journal} {Nat. Phys.}\ }\textbf {\bibinfo {volume} {21}},\ \bibinfo {pages} {25} (\bibinfo {year} {2025}{\natexlab{a}})}\BibitemShut {NoStop}%
\bibitem [{\citenamefont {Halimeh}\ \emph {et~al.}(2025{\natexlab{b}})\citenamefont {Halimeh}, \citenamefont {Mueller}, \citenamefont {Knolle}, \citenamefont {Papi{\'c}},\ and\ \citenamefont {Davoudi}}]{halimeh2025quantumsimulationoutofequilibriumdynamics}%
  \BibitemOpen
  \bibfield  {author} {\bibinfo {author} {\bibfnamefont {J.~C.}\ \bibnamefont {Halimeh}}, \bibinfo {author} {\bibfnamefont {N.}~\bibnamefont {Mueller}}, \bibinfo {author} {\bibfnamefont {J.}~\bibnamefont {Knolle}}, \bibinfo {author} {\bibfnamefont {Z.}~\bibnamefont {Papi{\'c}}},\ and\ \bibinfo {author} {\bibfnamefont {Z.}~\bibnamefont {Davoudi}},\ }\href {https://arxiv.org/abs/2509.03586} {\bibinfo {title} {Quantum simulation of out-of-equilibrium dynamics in gauge theories}} (\bibinfo {year} {2025}{\natexlab{b}}),\ \Eprint {https://arxiv.org/abs/2509.03586} {arXiv:2509.03586 [quant-ph]} \BibitemShut {NoStop}%
\bibitem [{\citenamefont {Zohar}\ \emph {et~al.}(2017)\citenamefont {Zohar}, \citenamefont {Farace}, \citenamefont {Reznik},\ and\ \citenamefont {Cirac}}]{zohar2017digital}%
  \BibitemOpen
  \bibfield  {author} {\bibinfo {author} {\bibfnamefont {E.}~\bibnamefont {Zohar}}, \bibinfo {author} {\bibfnamefont {A.}~\bibnamefont {Farace}}, \bibinfo {author} {\bibfnamefont {B.}~\bibnamefont {Reznik}},\ and\ \bibinfo {author} {\bibfnamefont {J.~I.}\ \bibnamefont {Cirac}},\ }\bibfield  {title} {\bibinfo {title} {Digital quantum simulation of $\mathbb{{Z}}_{2}$ lattice gauge theories with dynamical fermionic matter},\ }\href {https://doi.org/10.1103/PhysRevLett.118.070501} {\bibfield  {journal} {\bibinfo  {journal} {Phys. Rev. Lett.}\ }\textbf {\bibinfo {volume} {118}},\ \bibinfo {pages} {070501} (\bibinfo {year} {2017})}\BibitemShut {NoStop}%
\bibitem [{\citenamefont {Barbiero}\ \emph {et~al.}(2019)\citenamefont {Barbiero}, \citenamefont {Schweizer}, \citenamefont {Aidelsburger}, \citenamefont {Demler}, \citenamefont {Goldman},\ and\ \citenamefont {Grusdt}}]{barbiero2019coupling}%
  \BibitemOpen
  \bibfield  {author} {\bibinfo {author} {\bibfnamefont {L.}~\bibnamefont {Barbiero}}, \bibinfo {author} {\bibfnamefont {C.}~\bibnamefont {Schweizer}}, \bibinfo {author} {\bibfnamefont {M.}~\bibnamefont {Aidelsburger}}, \bibinfo {author} {\bibfnamefont {E.}~\bibnamefont {Demler}}, \bibinfo {author} {\bibfnamefont {N.}~\bibnamefont {Goldman}},\ and\ \bibinfo {author} {\bibfnamefont {F.}~\bibnamefont {Grusdt}},\ }\bibfield  {title} {\bibinfo {title} {Coupling ultracold matter to dynamical gauge fields in optical lattices: From flux attachment to $\mathbb{{Z}}_2$ lattice gauge theories},\ }\href {https://doi.org/10.1126/sciadv.aav7444} {\bibfield  {journal} {\bibinfo  {journal} {Sci. Adv.}\ }\textbf {\bibinfo {volume} {5}},\ \bibinfo {pages} {eaav7444} (\bibinfo {year} {2019})} \BibitemShut {NoStop}%
\bibitem [{\citenamefont {Lumia}\ \emph {et~al.}(2022)\citenamefont {Lumia}, \citenamefont {Torta}, \citenamefont {Mbeng}, \citenamefont {Santoro}, \citenamefont {Ercolessi}, \citenamefont {Burrello},\ and\ \citenamefont {Wauters}}]{lumia2022}%
  \BibitemOpen
  \bibfield  {author} {\bibinfo {author} {\bibfnamefont {L.}~\bibnamefont {Lumia}}, \bibinfo {author} {\bibfnamefont {P.}~\bibnamefont {Torta}}, \bibinfo {author} {\bibfnamefont {G.~B.}\ \bibnamefont {Mbeng}}, \bibinfo {author} {\bibfnamefont {G.~E.}\ \bibnamefont {Santoro}}, \bibinfo {author} {\bibfnamefont {E.}~\bibnamefont {Ercolessi}}, \bibinfo {author} {\bibfnamefont {M.}~\bibnamefont {Burrello}},\ and\ \bibinfo {author} {\bibfnamefont {M.~M.}\ \bibnamefont {Wauters}},\ }\bibfield  {title} {\bibinfo {title} {Two-dimensional $\mathbb{{Z}}_{2}$ lattice gauge theory on a near-term quantum simulator: Variational quantum optimization, confinement, and topological order},\ }\href {https://doi.org/10.1103/PRXQuantum.3.020320} {\bibfield  {journal} {\bibinfo  {journal} {PRX Quantum}\ }\textbf {\bibinfo {volume} {3}},\ \bibinfo {pages} {020320} (\bibinfo {year} {2022})}\BibitemShut {NoStop}%
\bibitem [{\citenamefont {Alexandrou}\ \emph {et~al.}(2025)\citenamefont {Alexandrou}, \citenamefont {Athenodorou}, \citenamefont {Blekos}, \citenamefont {Polykratis},\ and\ \citenamefont {K\"uhn}}]{alexandrou2025}%
  \BibitemOpen
  \bibfield  {author} {\bibinfo {author} {\bibfnamefont {C.}~\bibnamefont {Alexandrou}}, \bibinfo {author} {\bibfnamefont {A.}~\bibnamefont {Athenodorou}}, \bibinfo {author} {\bibfnamefont {K.}~\bibnamefont {Blekos}}, \bibinfo {author} {\bibfnamefont {G.}~\bibnamefont {Polykratis}},\ and\ \bibinfo {author} {\bibfnamefont {S.}~\bibnamefont {K\"uhn}},\ }\bibfield  {title} {\bibinfo {title} {Realizing string breaking dynamics in a $\mathbb{{Z}}_{2}$ lattice gauge theory on quantum hardware},\ }\href {https://doi.org/10.1103/r6sr-dv13} {\bibfield  {journal} {\bibinfo  {journal} {Phys. Rev. D}\ }\textbf {\bibinfo {volume} {112}},\ \bibinfo {pages} {114506} (\bibinfo {year} {2025})}\BibitemShut {NoStop}%
\bibitem [{\citenamefont {Cochran}\ \emph {et~al.}(2025)\citenamefont {Cochran} \emph {et~al.}}]{Cochran2025}%
  \BibitemOpen
  \bibfield  {author} {\bibinfo {author} {\bibfnamefont {T.~A.}\ \bibnamefont {Cochran}} \emph {et~al.},\ }\bibfield  {title} {\bibinfo {title} {Visualizing dynamics of charges and strings in (2+1)d lattice gauge theories},\ }\href {https://doi.org/10.1038/s41586-025-08999-9} {\bibfield  {journal} {\bibinfo  {journal} {Nature}\ }\textbf {\bibinfo {volume} {642}},\ \bibinfo {pages} {315} (\bibinfo {year} {2025})}\BibitemShut {NoStop}%
\bibitem [{\citenamefont {Borla}\ \emph {et~al.}(2025)\citenamefont {Borla}, \citenamefont {Osborne}, \citenamefont {Moroz},\ and\ \citenamefont {Halimeh}}]{borla2025stringbreaking}%
  \BibitemOpen
  \bibfield  {author} {\bibinfo {author} {\bibfnamefont {U.}~\bibnamefont {Borla}}, \bibinfo {author} {\bibfnamefont {J.~J.}\ \bibnamefont {Osborne}}, \bibinfo {author} {\bibfnamefont {S.}~\bibnamefont {Moroz}},\ and\ \bibinfo {author} {\bibfnamefont {J.~C.}\ \bibnamefont {Halimeh}},\ }\href {https://arxiv.org/abs/2501.17929} {\bibinfo {title} {String breaking in a $2+1$d $\mathbb{Z}_2$ lattice gauge theory}} (\bibinfo {year} {2025}),\ \Eprint {https://arxiv.org/abs/2501.17929} {arXiv:2501.17929 [quant-ph]} \BibitemShut {NoStop}%
\bibitem [{\citenamefont {Cobos}\ \emph {et~al.}(2025)\citenamefont {Cobos}, \citenamefont {Fraxanet}, \citenamefont {Benito}, \citenamefont {di~Marcantonio}, \citenamefont {Rivero}, \citenamefont {Kap{\'a}s}, \citenamefont {Werner}, \citenamefont {Legeza}, \citenamefont {Bermudez},\ and\ \citenamefont {Rico}}]{cobos2025realtimedynamics21dgauge}%
  \BibitemOpen
  \bibfield  {author} {\bibinfo {author} {\bibfnamefont {J.}~\bibnamefont {Cobos}}, \bibinfo {author} {\bibfnamefont {J.}~\bibnamefont {Fraxanet}}, \bibinfo {author} {\bibfnamefont {C.}~\bibnamefont {Benito}}, \bibinfo {author} {\bibfnamefont {F.}~\bibnamefont {di~Marcantonio}}, \bibinfo {author} {\bibfnamefont {P.}~\bibnamefont {Rivero}}, \bibinfo {author} {\bibfnamefont {K.}~\bibnamefont {Kap{\'a}s}}, \bibinfo {author} {\bibfnamefont {M.~A.}\ \bibnamefont {Werner}}, \bibinfo {author} {\bibfnamefont {\"O.}~\bibnamefont {Legeza}}, \bibinfo {author} {\bibfnamefont {A.}~\bibnamefont {Bermudez}},\ and\ \bibinfo {author} {\bibfnamefont {E.}~\bibnamefont {Rico}},\ }\href {https://arxiv.org/abs/2507.08088} {\bibinfo {title} {Real-time dynamics in a (2+1)-d gauge theory: The stringy nature on a superconducting quantum simulator}} (\bibinfo {year} {2025}),\ \Eprint {https://arxiv.org/abs/2507.08088} {arXiv:2507.08088 [quant-ph]} \BibitemShut {NoStop}%
\bibitem [{\citenamefont {Xu}\ \emph {et~al.}(2025)\citenamefont {Xu}, \citenamefont {Borla}, \citenamefont {Moroz},\ and\ \citenamefont {Halimeh}}]{Xu2025StringBreakingGlueball}%
  \BibitemOpen
  \bibfield  {author} {\bibinfo {author} {\bibfnamefont {K.}~\bibnamefont {Xu}}, \bibinfo {author} {\bibfnamefont {U.}~\bibnamefont {Borla}}, \bibinfo {author} {\bibfnamefont {S.}~\bibnamefont {Moroz}},\ and\ \bibinfo {author} {\bibfnamefont {J.~C.}\ \bibnamefont {Halimeh}},\ }\href {https://arxiv.org/abs/2507.01950} {\bibinfo {title} {String breaking dynamics and glueball formation in a $2+1$d lattice gauge theory}} (\bibinfo {year} {2025}),\ \Eprint {https://arxiv.org/abs/2507.01950} {arXiv:2507.01950 [quant-ph]} \BibitemShut {NoStop}%
\bibitem [{\citenamefont {Xu}\ \emph {et~al.}(2026)\citenamefont {Xu}, \citenamefont {Borla}, \citenamefont {Hemery}, \citenamefont {Joshi}, \citenamefont {Dreyer}, \citenamefont {Rinaldi},\ and\ \citenamefont {Halimeh}}]{xu2026observationglueballexcitationsstring}%
  \BibitemOpen
  \bibfield  {author} {\bibinfo {author} {\bibfnamefont {K.}~\bibnamefont {Xu}}, \bibinfo {author} {\bibfnamefont {U.}~\bibnamefont {Borla}}, \bibinfo {author} {\bibfnamefont {K.}~\bibnamefont {Hemery}}, \bibinfo {author} {\bibfnamefont {R.}~\bibnamefont {Joshi}}, \bibinfo {author} {\bibfnamefont {H.}~\bibnamefont {Dreyer}}, \bibinfo {author} {\bibfnamefont {E.}~\bibnamefont {Rinaldi}},\ and\ \bibinfo {author} {\bibfnamefont {J.~C.}\ \bibnamefont {Halimeh}},\ }\href {https://arxiv.org/abs/2604.07435} {\bibinfo {title} {Observation of glueball excitations and string breaking in a $2+1$d $\mathbb{Z}_2$ lattice gauge theory on a trapped-ion quantum computer}} (\bibinfo {year} {2026}),\ \Eprint {https://arxiv.org/abs/2604.07435} {arXiv:2604.07435 [hep-lat]} \BibitemShut {NoStop}%
\bibitem [{\citenamefont {{Google Quantum AI and Collaborators}}(2026)}]{googledfl2026}%
  \BibitemOpen
  \bibfield  {author} {\bibinfo {author} {\bibnamefont {{Google Quantum AI and Collaborators}}},\ }\bibfield  {title} {\bibinfo {title} {Observation of disorder-free localization using a (2+1)d lattice gauge theory on a quantum processor},\ }\href {https://doi.org/10.1126/science.adr9680} {\bibfield  {journal}
  {\bibinfo  {journal} {Science}\ }\textbf {\bibinfo {volume} {393}},\ \bibinfo {pages} {71} (\bibinfo {year} {2026})} \BibitemShut {NoStop}%
\bibitem [{\citenamefont {Homeier}\ \emph {et~al.}(2021)\citenamefont {Homeier}, \citenamefont {Schweizer}, \citenamefont {Aidelsburger}, \citenamefont {Fedorov},\ and\ \citenamefont {Grusdt}}]{homeier2021}%
  \BibitemOpen
  \bibfield  {author} {\bibinfo {author} {\bibfnamefont {L.}~\bibnamefont {Homeier}}, \bibinfo {author} {\bibfnamefont {C.}~\bibnamefont {Schweizer}}, \bibinfo {author} {\bibfnamefont {M.}~\bibnamefont {Aidelsburger}}, \bibinfo {author} {\bibfnamefont {A.}~\bibnamefont {Fedorov}},\ and\ \bibinfo {author} {\bibfnamefont {F.}~\bibnamefont {Grusdt}},\ }\bibfield  {title} {\bibinfo {title} {$\mathbb{{Z}}_{2}$ lattice gauge theories and {K}itaev's toric code: A scheme for analog quantum simulation},\ }\href {https://doi.org/10.1103/PhysRevB.104.085138} {\bibfield  {journal} {\bibinfo  {journal} {Phys. Rev. B}\ }\textbf {\bibinfo {volume} {104}},\ \bibinfo {pages} {085138} (\bibinfo {year} {2021})}\BibitemShut {NoStop}%
\bibitem [{\citenamefont {Bravyi}\ and\ \citenamefont {Kitaev}(2002)}]{BRAVYI2002210}%
  \BibitemOpen
  \bibfield  {author} {\bibinfo {author} {\bibfnamefont {S.~B.}\ \bibnamefont {Bravyi}}\ and\ \bibinfo {author} {\bibfnamefont {A.~Y.}\ \bibnamefont {Kitaev}},\ }\bibfield  {title} {\bibinfo {title} {Fermionic quantum computation},\ }\href {https://doi.org/10.1006/aphy.2002.6254} {\bibfield  {journal} {\bibinfo  {journal} {Ann. Phys.}\ }\textbf {\bibinfo {volume} {298}},\ \bibinfo {pages} {210} (\bibinfo {year} {2002})}\BibitemShut {NoStop}%
\bibitem [{\citenamefont {Derby}\ \emph {et~al.}(2021)\citenamefont {Derby}, \citenamefont {Klassen}, \citenamefont {Bausch},\ and\ \citenamefont {Cubitt}}]{derby2021}%
  \BibitemOpen
  \bibfield  {author} {\bibinfo {author} {\bibfnamefont {C.}~\bibnamefont {Derby}}, \bibinfo {author} {\bibfnamefont {J.}~\bibnamefont {Klassen}}, \bibinfo {author} {\bibfnamefont {J.}~\bibnamefont {Bausch}},\ and\ \bibinfo {author} {\bibfnamefont {T.}~\bibnamefont {Cubitt}},\ }\bibfield  {title} {\bibinfo {title} {Compact fermion to qubit mappings},\ }\href {https://doi.org/10.1103/PhysRevB.104.035118} {\bibfield  {journal} {\bibinfo  {journal} {Phys. Rev. B}\ }\textbf {\bibinfo {volume} {104}},\ \bibinfo {pages} {035118} (\bibinfo {year} {2021})}\BibitemShut {NoStop}%
\bibitem [{\citenamefont {Wilczek}(1982)}]{wilczek1982}%
  \BibitemOpen
  \bibfield  {author} {\bibinfo {author} {\bibfnamefont {F.}~\bibnamefont {Wilczek}},\ }\bibfield  {title} {\bibinfo {title} {Magnetic flux, angular momentum, and statistics},\ }\href {https://doi.org/10.1103/PhysRevLett.48.1144} {\bibfield  {journal} {\bibinfo  {journal} {Phys. Rev. Lett.}\ }\textbf {\bibinfo {volume} {48}},\ \bibinfo {pages} {1144} (\bibinfo {year} {1982})}\BibitemShut {NoStop}%
\bibitem [{\citenamefont {Eliezer}\ and\ \citenamefont {Semenoff}(1992)}]{ELIEZER199266}%
  \BibitemOpen
  \bibfield  {author} {\bibinfo {author} {\bibfnamefont {D.}~\bibnamefont {Eliezer}}\ and\ \bibinfo {author} {\bibfnamefont {G.}~\bibnamefont {Semenoff}},\ }\bibfield  {title} {\bibinfo {title} {Anyonization of lattice {C}hern--{S}imons theory},\ }\href {https://doi.org/10.1016/0003-4916(92)90339-N} {\bibfield  {journal} {\bibinfo  {journal} {Ann. Phys.}\ }\textbf {\bibinfo {volume} {217}},\ \bibinfo {pages} {66} (\bibinfo {year} {1992})}\BibitemShut {NoStop}%
\bibitem [{\citenamefont {Peng}\ \emph {et~al.}(2026)\citenamefont {Peng}, \citenamefont {Diamantini}, \citenamefont {Funcke}, \citenamefont {Hassan}, \citenamefont {Jansen}, \citenamefont {K\"uhn}, \citenamefont {Luo},\ and\ \citenamefont {Naredi}}]{peng2026}%
  \BibitemOpen
  \bibfield  {author} {\bibinfo {author} {\bibfnamefont {C.}~\bibnamefont {Peng}}, \bibinfo {author} {\bibfnamefont {M.~C.}\ \bibnamefont {Diamantini}}, \bibinfo {author} {\bibfnamefont {L.}~\bibnamefont {Funcke}}, \bibinfo {author} {\bibfnamefont {S.~M.~A.}\ \bibnamefont {Hassan}}, \bibinfo {author} {\bibfnamefont {K.}~\bibnamefont {Jansen}}, \bibinfo {author} {\bibfnamefont {S.}~\bibnamefont {K\"uhn}}, \bibinfo {author} {\bibfnamefont {D.}~\bibnamefont {Luo}},\ and\ \bibinfo {author} {\bibfnamefont {P.}~\bibnamefont {Naredi}},\ }\bibfield  {title} {\bibinfo {title} {Hamiltonian lattice formulation of compact {M}axwell--{C}hern--{S}imons theory},\ }\href {https://doi.org/10.1103/rrzw-s456} {\bibfield  {journal} {\bibinfo  {journal} {Phys. Rev. D}\ }\textbf {\bibinfo {volume} {114}},\ \bibinfo {pages} {034511} (\bibinfo {year} {2026})}\BibitemShut {NoStop}%
\bibitem [{\citenamefont {Verstraete}\ \emph {et~al.}(2008)\citenamefont {Verstraete}, \citenamefont {Murg},\ and\ \citenamefont {Cirac}}]{Verstraete01032008}%
  \BibitemOpen
  \bibfield  {author} {\bibinfo {author} {\bibfnamefont {F.}~\bibnamefont {Verstraete}}, \bibinfo {author} {\bibfnamefont {V.}~\bibnamefont {Murg}},\ and\ \bibinfo {author} {\bibfnamefont {J.~I.}\ \bibnamefont {Cirac}},\ }\bibfield  {title} {\bibinfo {title} {Matrix product states, projected entangled pair states, and variational renormalization group methods for quantum spin systems},\ }\href {https://doi.org/10.1080/14789940801912366} {\bibfield  {journal} {\bibinfo  {journal} {Adv. Phys.}\ }\textbf {\bibinfo {volume} {57}},\ \bibinfo {pages} {143} (\bibinfo {year} {2008})} \BibitemShut {NoStop}%
\bibitem [{\citenamefont {Naumann}\ \emph {et~al.}(2024)\citenamefont {Naumann}, \citenamefont {Weerda}, \citenamefont {Rizzi}, \citenamefont {Eisert},\ and\ \citenamefont {Schmoll}}]{naumann2024peps}%
  \BibitemOpen
  \bibfield  {author} {\bibinfo {author} {\bibfnamefont {J.}~\bibnamefont {Naumann}}, \bibinfo {author} {\bibfnamefont {E.~L.}\ \bibnamefont {Weerda}}, \bibinfo {author} {\bibfnamefont {M.}~\bibnamefont {Rizzi}}, \bibinfo {author} {\bibfnamefont {J.}~\bibnamefont {Eisert}},\ and\ \bibinfo {author} {\bibfnamefont {P.}~\bibnamefont {Schmoll}},\ }\bibfield  {title} {\bibinfo {title} {An introduction to infinite projected entangled-pair state methods for variational ground state simulations using automatic differentiation},\ }\href {https://doi.org/10.21468/SciPostPhysLectNotes.86} {\bibfield  {journal} {\bibinfo  {journal} {SciPost Phys. Lect. Notes}\ }\textbf {\bibinfo {volume} {86}} (\bibinfo {year} {2024})}\BibitemShut {NoStop}%
\bibitem [{\citenamefont {Carleo}\ and\ \citenamefont {Troyer}(2017)}]{Carleo_2017}%
  \BibitemOpen
  \bibfield  {author} {\bibinfo {author} {\bibfnamefont {G.}~\bibnamefont {Carleo}}\ and\ \bibinfo {author} {\bibfnamefont {M.}~\bibnamefont {Troyer}},\ }\bibfield  {title} {\bibinfo {title} {Solving the quantum many-body problem with artificial neural networks},\ }\href {https://doi.org/10.1126/science.aag2302} {\bibfield  {journal} {\bibinfo  {journal} {Science}\ }\textbf {\bibinfo {volume} {355}},\ \bibinfo {pages} {602} (\bibinfo {year} {2017})}\BibitemShut {NoStop}%
\bibitem [{\citenamefont {Lange}\ \emph {et~al.}(2024)\citenamefont {Lange}, \citenamefont {Van~de Walle}, \citenamefont {Abedinnia},\ and\ \citenamefont {Bohrdt}}]{Lange_2024}%
  \BibitemOpen
  \bibfield  {author} {\bibinfo {author} {\bibfnamefont {H.}~\bibnamefont {Lange}}, \bibinfo {author} {\bibfnamefont {A.}~\bibnamefont {Van~de Walle}}, \bibinfo {author} {\bibfnamefont {A.}~\bibnamefont {Abedinnia}},\ and\ \bibinfo {author} {\bibfnamefont {A.}~\bibnamefont {Bohrdt}},\ }\bibfield  {title} {\bibinfo {title} {From architectures to applications: a review of neural quantum states},\ }\href {https://doi.org/10.1088/2058-9565/ad7168} {\bibfield  {journal} {\bibinfo  {journal} {Quantum Sci. Technol.}\ }\textbf {\bibinfo {volume} {9}},\ \bibinfo {pages} {040501} (\bibinfo {year} {2024})}\BibitemShut {NoStop}%
\bibitem [{\citenamefont {Kufel}\ \emph {et~al.}(2025)\citenamefont {Kufel}, \citenamefont {Kemp}, \citenamefont {Vu}, \citenamefont {Linsel}, \citenamefont {Laumann},\ and\ \citenamefont {Yao}}]{kufel2025}%
  \BibitemOpen
  \bibfield  {author} {\bibinfo {author} {\bibfnamefont {D.~S.}\ \bibnamefont {Kufel}}, \bibinfo {author} {\bibfnamefont {J.}~\bibnamefont {Kemp}}, \bibinfo {author} {\bibfnamefont {D.}~\bibnamefont {Vu}}, \bibinfo {author} {\bibfnamefont {S.~M.}\ \bibnamefont {Linsel}}, \bibinfo {author} {\bibfnamefont {C.~R.}\ \bibnamefont {Laumann}},\ and\ \bibinfo {author} {\bibfnamefont {N.~Y.}\ \bibnamefont {Yao}},\ }\bibfield  {title} {\bibinfo {title} {Approximately symmetric neural networks for quantum spin liquids},\ }\href {https://doi.org/10.1103/pgnx-11ph} {\bibfield  {journal} {\bibinfo  {journal} {Phys. Rev. Lett.}\ }\textbf {\bibinfo {volume} {135}},\ \bibinfo {pages} {056702} (\bibinfo {year} {2025})}\BibitemShut {NoStop}%
\bibitem [{\citenamefont {Hauschild}\ \emph {et~al.}(2024)\citenamefont {Hauschild}, \citenamefont {Unfried}, \citenamefont {Anand}, \citenamefont {Andrews}, \citenamefont {Bintz}, \citenamefont {Borla}, \citenamefont {Divic}, \citenamefont {Drescher}, \citenamefont {Geiger}, \citenamefont {Hefel}, \citenamefont {H{\'e}mery}, \citenamefont {Kadow}, \citenamefont {Kemp}, \citenamefont {Kirchner}, \citenamefont {Liu}, \citenamefont {M{\"o}ller}, \citenamefont {Parker}, \citenamefont {Rader}, \citenamefont {Romen}, \citenamefont {Scalet}, \citenamefont {Schoonderwoerd}, \citenamefont {Schulz}, \citenamefont {Soejima}, \citenamefont {Thoma}, \citenamefont {Wu}, \citenamefont {Zechmann}, \citenamefont {Zweng}, \citenamefont {Mong}, \citenamefont {Zaletel},\ and\ \citenamefont {Pollmann}}]{tenpy2024}%
  \BibitemOpen
  \bibfield  {author} {\bibinfo {author} {\bibfnamefont {J.}~\bibnamefont {Hauschild}}, \bibinfo {author} {\bibfnamefont {J.}~\bibnamefont {Unfried}}, \bibinfo {author} {\bibfnamefont {S.}~\bibnamefont {Anand}}, \bibinfo {author} {\bibfnamefont {B.}~\bibnamefont {Andrews}}, \bibinfo {author} {\bibfnamefont {M.}~\bibnamefont {Bintz}}, \bibinfo {author} {\bibfnamefont {U.}~\bibnamefont {Borla}}, \bibinfo {author} {\bibfnamefont {S.}~\bibnamefont {Divic}}, \bibinfo {author} {\bibfnamefont {M.}~\bibnamefont {Drescher}}, \bibinfo {author} {\bibfnamefont {J.}~\bibnamefont {Geiger}}, \bibinfo {author} {\bibfnamefont {M.}~\bibnamefont {Hefel}}, \bibinfo {author} {\bibfnamefont {K.}~\bibnamefont {H{\'e}mery}}, \bibinfo {author} {\bibfnamefont {W.}~\bibnamefont {Kadow}}, \bibinfo {author} {\bibfnamefont {J.}~\bibnamefont {Kemp}}, \bibinfo {author} {\bibfnamefont {N.}~\bibnamefont {Kirchner}}, \bibinfo {author} {\bibfnamefont {V.~S.}\ \bibnamefont {Liu}}, \bibinfo {author} {\bibfnamefont {G.}~\bibnamefont {M{\"o}ller}},
  \bibinfo {author} {\bibfnamefont {D.}~\bibnamefont {Parker}}, \bibinfo {author} {\bibfnamefont {M.}~\bibnamefont {Rader}}, \bibinfo {author} {\bibfnamefont {A.}~\bibnamefont {Romen}}, \bibinfo {author} {\bibfnamefont {S.}~\bibnamefont {Scalet}}, \bibinfo {author} {\bibfnamefont {L.}~\bibnamefont {Schoonderwoerd}}, \bibinfo {author} {\bibfnamefont {M.}~\bibnamefont {Schulz}}, \bibinfo {author} {\bibfnamefont {T.}~\bibnamefont {Soejima}}, \bibinfo {author} {\bibfnamefont {P.}~\bibnamefont {Thoma}}, \bibinfo {author} {\bibfnamefont {Y.}~\bibnamefont {Wu}}, \bibinfo {author} {\bibfnamefont {P.}~\bibnamefont {Zechmann}}, \bibinfo {author} {\bibfnamefont {L.}~\bibnamefont {Zweng}}, \bibinfo {author} {\bibfnamefont {R.~S.~K.}\ \bibnamefont {Mong}}, \bibinfo {author} {\bibfnamefont {M.~P.}\ \bibnamefont {Zaletel}},\ and\ \bibinfo {author} {\bibfnamefont {F.}~\bibnamefont {Pollmann}},\ }\bibfield  {title} {\bibinfo {title} {{Tensor network Python (TeNPy) version 1}},\ }\href
  {https://doi.org/10.21468/SciPostPhysCodeb.41} {\bibfield  {journal} {\bibinfo  {journal} {SciPost Phys. Codebases}\ }\textbf {\bibinfo {volume} {41}} (\bibinfo {year} {2024})}\BibitemShut {NoStop}%
\bibitem [{\citenamefont {McCulloch}\ and\ \citenamefont {Osborne}()}]{mptoolkit}%
  \BibitemOpen
  \bibfield  {author} {\bibinfo {author} {\bibfnamefont {I.~P.}\ \bibnamefont {McCulloch}}\ and\ \bibinfo {author} {\bibfnamefont {J.~J.}\ \bibnamefont {Osborne}},\ }\href@noop {} {\bibinfo {title} {Matrix product toolkit}},\ \bibinfo {howpublished} {\url{https://github.com/mptoolkit}}\BibitemShut {NoStop}%
\end{thebibliography}
\end{document}